\documentclass[useAMS]{mn2e}

\usepackage{graphicx}
\usepackage{times}
\newcommand{\Teff}{\ensuremath{T_{\rm eff}}}                % Effective temperature symbol
\newcommand{\logg}{\ensuremath{\log g}}                     % log(g) symbol
\newcommand{\Msun}{\ensuremath{\,{\rm M}_\odot}}            % Solar mass symbol
\newcommand{\Rsun}{\ensuremath{\,{\rm R}_\odot}}            % Solar radius symbol
\newcommand{\Lsun}{\ensuremath{\,{\rm L}_\odot}}            % Solar luminosity symbol
\newcommand{\ion}[2]{{#1}\,{\sc {\small{#2}}}}              % This creates ion symbols using small caps
\newcommand{\epshe}{\ensuremath{\epsilon(\rm He)}}          % Helium abundance symbol
\newcommand{\kms}{\,km\,s$^{-1}$}                           % km/s symbol
\newcommand{\cmss}{\,cm\,s$^{-2}$}                          % cm/s/2 symbol
\newcommand{\micro}{\ensuremath{v_{\rm turb}}}              % microturbulence velocity
\newcommand{\Veq}{\ensuremath{V_{\rm eq}}}                  % Stellar equatorial rotational velocity symbol
\newcommand{\Vsync}{\ensuremath{V_{\rm synch}}}             % Synchronous rotational velocity symbol
\newcommand{\EBV}{\ensuremath{E_{B-V}}}                     % E(B-V) symbol
\newcommand{\spd}{{\sc spd}}
\newcommand{\mc}[1]{\multicolumn{2}{c}{#1}}

\title[Chemical evolution of high-mass stars. II.]
    {Chemical evolution of high-mass stars in close binaries. II. The evolved component
 of the eclipsing binary V380\,Cygni}

\author[K.\ Pavlovski et al.]
       {K.\ Pavlovski$^1$, E.\ Tamajo$^1$, P.\ Koubsk\'{y}$^2$, J.\ Southworth$^3$, S.\ Yang$^4$ and  V.\ Kolbas$^1$ \\
       $^1$\,Department of Physics, University of Zagreb, Bijeni\v{c}ka cesta 32, 10000 Zagreb, Croatia \\
       $^2$\,Astronomical Institute of the Academy of Sciences, 251\,65 Ond\v{r}ejov, Czech Republic \\
       $^3$\,Department of Physics, University of Warwick, Coventry CV4 7AL, UK \\
       $^4$\,Department of Physics and Astronomy, University of Victoria, Victoria, BC V8W3P6, Canada}

\date{}

\pagerange{\pageref{firstpage}--\pageref{lastpage}} \pubyear{2009}

\begin{document} \maketitle \label{firstpage} %%%%%%%%%%%%%%%%%%%%%%%%%%%%%%%%%%%%%%%%%%%%%%%%%%%%%%%%%%%%%%%%%%%%
%%%%%%%%%%%%%%%%%%%%%%%%%%%%%%%%%%%%%%%%%%%%%%%%%%%%%%%%%%%%%%%%%%%%%%%%%%%%%%%%%%%%%%%%%%%%%%%%%%%%%%%%%%%%%%%%%%

\begin{abstract}
The eclipsing and double-lined spectroscopic binary V380\,Cyg is an extremely important probe of stellar
evolution: its primary component is a high-mass star at the brink of leaving the main sequence whereas
the secondary star is still in the early part of its main sequence lifetime.
We present extensive high-resolution \'echelle and grating spectroscopy from Ond\v{r}ejov, Calar Alto,
Victoria and La Palma. We apply spectral disentangling to unveil the individual spectra of the
two stars and obtain new spectroscopic elements. The secondary star contributes only about 6\% of
the total light, which remains the main limitation to measuring the system's characteristics.
We determine improved physical properties, finding masses $13.1 \pm 0.3$ and $7.8 \pm 0.1$ \Msun,
radii $16.2 \pm 0.3$ and $4.06 \pm 0.08$ \Rsun, and effective temperatures $21\,750 \pm 280$
and $21\,600 \pm 550$ K, for the primary and secondary components respectively. We perform a detailed 
abundance analysis by fitting non-LTE theoretical line profiles to the disentangled spectrum of the 
evolved primary star, and reveal an elemental abundance pattern reminiscent of a typical nearby B star. 
Contrary to the predictions of recent theoretical evolution models with rotational mixing, no trace of 
abundance modifications due to the CNO cycle are detected. No match can be found between the 
predictions of these models and the properties of the primary star: a mass discrepancy of 
1.5\Msun\ exists and remains unexplained.
\end{abstract}


%%%%%%%%%%%%%%%%%%%%%%%%%%%%%%%%%%%%%%%%%%%%%%%%%%%%%%%%%%%%%%%%%%%%%%%%%%%%%%%%%%%%%%%%%%%%%%%%%%%%%%%%%%%%%%%%%%

\section{Introduction}                                                                           \label{sec:intro}

The last decade has witnessed a huge improvement in modelling the structure and evolution of stars, 
particularly for higher masses. The inclusion of effects such as rotation and/or magnetic fields have 
caused substantial changes in the resulting predictions (see Maeder \& Meynet 2000; Langer et al.\ 2008). 
Some of these concern evolutionary changes in the chemical composition of stellar atmospheres. Due to 
the CNO cycle in the core of high-mass stars some elements are enhanced, such as helium and nitrogen, 
and some are depleted, like carbon and to a lesser extent oxygen. With deep mixing due to rotation, 
the products of core nucleosynthesis are predicted to be brought to the stellar surface and cause 
changes in atmospheric abundance patterns. Rotational mixing is so efficient that changes in the 
atmospheric composition should be identifiable whilst the star is still on the main sequence (MS).

Anomalous abundances of helium and CNO elements have been noticed for many years (Leushin 1988, 
Lyubimkov 1998), and confirmed in detail by Gies \& Lambert (1992), and more recently for helium 
abundances in B stars by Lyubimkov et al.\ (2004) and Huang \& Gies (2006). However, more ambitious 
observational studies indicate that the situation is not straightforward. In a survey of OB stars 
in our Galaxy and the Magellanic Clouds, multi-object spectroscopy for about 750 stars were secured 
by Evans et al.\ (2005, 2006). Elemental abundances were separately determined for both slow 
(Trundle et al.\ 2007, Hunter et al.\ 2007) and fast (Hunter et al.\ 2008, 2009) rotators in 
the LMC. Since theoretical calculations predict the strongest effect to be the enhancement of 
the nitrogen abundance (about 0.5 dex up to the terminal age of the main sequence; TAMS), 
Hunter et al.\ (2008) examined the abundances of this element in detail. They found a rather 
complex behaviour: nitrogen enrichment is correlated with projected rotational velocity, but 
there are some slow rotators with a high nitrogen abundance. Their findings corroborate a recent 
study by Morel et al.\ (2006), who analysed a sample of slowly rotating Galactic $\beta$ Cephei 
stars and found a group with nitrogen enrichment. Importantly, Morel et al.\ (2006) were able to 
link this with enhanced magnetic field strengths. It is clear that several physical phenomena and 
processes affect the surface chemical compositions of high-mass stars, and it is a challenge to 
disentangle them.

Empirical constraints on these processes remain hard to come by, despite a steady improvement in 
observational techniques and capabilities (see  Hilditch 2004). In this series of papers we aim 
to calibrate the abundance patterns and chemical evolution of high-mass stars by analysing detached 
eclipsing binaries (dEBs). These are vital for specifying empirical constraints on the properties 
of high-mass stars, since they are the primary source of directly measured stellar properties 
(Andersen 1991). Unfortunately, accurate (2\% or better) physical properties are available for 
only ten dEBs containing $>$10\Msun\ stars\footnote{An up-to-date compilation of the properties 
of well-studied dEBs is maintained at {\tt http://www.astro.keele.ac.uk/$\sim$jkt}}, and only four 
have observational constraints on their chemical composition. Chemical abundances are difficult to 
determine for high-mass dEBs for several reasons. Firstly, they tend to display only a small number 
of spectral lines. Secondly, their often high rotational velocities means the lines are wide and 
shallow. Thirdly, in dEBs the spectral lines from the two components interfere with each other 
(`line blending').

In a seminal work, Simon \& Sturm (1994) introduced the technique of {\em spectral disentangling} 
({\sc spd}), by which {\em individual} spectra of the component stars of double-lined spectroscopic 
binary systems can be deduced from observations covering a range of orbital phases. {\sc spd} can 
be used to measure spectroscopic orbits which are not affected by line blending (see Southworth \& 
Clausen 2007). The resulting disentangled spectra also have a much higher S/N than the original 
observations, so are well suited to chemical abundance analyses. As a bonus, the strong degeneracy 
between effective temperature (\Teff) and surface gravity (\logg) is not a problem for dEBs because
 surface gravities can be measured directly and to high accuracy (0.01\,dex or better). A detailed 
investigation of these possibilities is given by Pavlovski \& Hensberge (2005; hereafter PH05).

\subsection{The eclipsing binary system V380\,Cygni}                                         \label{sec:intro:v380}

V380\,Cyg is a very interesting probe of stellar structure and evolution because it contains a primary
 (star A) which is rather evolved ($\logg = 3.1$) and a less massive secondary (star B) which is not 
($\logg = 4.1$). A thorough study by Popper \& Guinan (1998) and Guinan et al.\ (2000, hereafter G2000) 
disclosed that star A is the more luminous and massive component of the system, lying near the 
`blue hook' of the MS, and that star B is a $\sim$3\,mag fainter unevolved MS star. Due to the 
faintness of star B, V380\,Cyg was for many years categorised as a single-lined spectroscopic 
binary (Batten 1962 and references therein). Hill \& Batten (1984) applied cross-correlation 
techniques to considerably improve the orbital elements. Further advances were possible only 
by high-resolution and high signal-to-noise (S/N) spectroscopy (Lyubimkov et al.\ 1996; Popper 
\& Guinan 1998).

The eclipses in V380\,Cyg are rather shallow (amplitudes of 0.12 and 0.09 mag) and occur on an
 orbital period of 12.4\,d. The radii of the stars are not determined to high accuracy despite 
considerable effort spent on securing a precise and complete light curve (G2000). These difficulties 
have contributed to disagreements over which of the two stars is hotter (Hill \& Batten 1984; 
Lyubimkov et al.\ 1996; G2000). The best estimate of ${\Teff}_{\rm A}$ to date came from fitting 
ultraviolet and visual spectrophotometry with model atmosphere energy distributions, giving 
$\Teff = 21\,350$\,K (G2000). There is also a discrepancy in the derived masses of the components. 
The two most recent studies gives $M_{\rm A} = 11.1 \pm 0.5$\Msun\ and $M_{\rm B} = 6.95 \pm 
0.25$\Msun\ (G2000) and $M_{\rm A} = 12.1 \pm 0.3$\Msun\ and $M_{\rm B} = 7.3 \pm 0.3$\Msun\ 
(Lyubimkov et al.\ 1996).

V380\,Cyg has an eccentric orbit and displays the phenomenon of apsidal motion with a rate of 
$\dot{\omega} = 24.0 \pm 1.8$ degrees per 100\,yr (G2000). To match the observed stellar properties 
to theoretical evolutionary tracks these authors found that a large core-overshooting parameter 
is required, $\alpha_{\rm ov} = 0.6 \pm 0.1$. In the HR diagram star A is predicted to be located 
near the blue point of the MS hook. G2000 drew a general conclusion that high-mass stars have large 
convective cores, so are more centrally condensed than predicted by standard evolutionary theory. 
However, theoretical work by Claret (2003, 2007) does not corroborate this conclusion, finding 
instead that $\alpha_{\rm ov} = 0.4^{+0.2}_{-0.3}$ for this dEB.

The chemical composition of V380\,Cyg, in particular the helium abundances, were studied by Leushin 
\& Topilskaya (1986) and Lyubimkov et al.\ (1996). Both groups found star A to be enriched in helium 
and star B to be normal. A major goal of our work is to improve and extend these results, and thus be 
able to perform a detailed comparison with theoretical stellar models.

%%%%%%%%%%%%%%%%%%%%%%%%%%%%%%%%%%%%%%%%%%%%%%%%%%%%%%%%%%%%%%%%%%%%%%%%%%%%%%%%%%%%%%%%%%%%%%%%%%%%%%%%%%%%%%%%%%

\section{Spectroscopic data}                                                                       \label{sec:obs}

Star B is barely visible in the optical spectrum of the binary -- G2000 estimated the light ratio in 
the visual band to be $L_{\rm A}/L_{\rm B} \sim 14.5$. A successful observational programme therefore 
requires high-resolution and high-S/N spectroscopy. For this study we have secured spectra at four 
different observatories, using either diffraction grating or \'echelle spectrographs. Observing logs 
are given in the Appendix (available in the electronic version of this work).

\subsection{Ond\v{r}ejov spectra}

59 spectra of V380\,Cyg were obtained at the Astronomical Institute of the Academy of Sciences of the 
Czech Republic in Ond\v{r}ejov. The observations were made in the years 2004 to 2007 with the 2\,m telescope 
and the coud\'e spectrograph. The spectral interval centred on the H$\alpha$ line covers about 500\,\AA\ 
at a dispersion of 17\,\AA/mm, whilst the spectral intervals centred on the H$\beta$ and H$\gamma$ lines 
cover around 250\,\AA, at a dispersion 8.5\,\AA/mm.

\subsection{Calar Alto spectra}

We obtained 43 spectra of V380\,Cyg in the course of two observing runs (May and August 2008) at the Centro 
Astron\'omico Hispano Alem\'an (CAHA) at Calar Alto, Spain. We used the 2.2\,m telescope, FOCES \'echelle 
spectrograph (Pfeiffer et al.\ 1998), and a Loral \#11i CCD binned 2$\times$2 to decrease the readout 
time. With a grating angle of 2724, prism angle of 130 and a 150\,$\mu$m slit we obtained a spectral 
coverage of roughly 3700--9200\,\AA\ in each exposure, at a resolving power of $R \approx 40\,000$. 
Wavelength calibration was performed using thorium-argon exposures, and flat-fields were obtained using
 a tungsten lamp. The observing conditions were generally good but several exposures were affected by clouds.

\subsection{Victoria spectra}

Spectra were obtained with the 1.2 m McKellar telescope and its coud\'{e} spectrograph at Dominion 
Astrophysical Observatory (DAO) with a dispersion of 9\,\AA/mm. A total of 27 spectra were secured in the 
region of the H$\alpha$ line ($\lambda\lambda$6150--6760), mostly in the years 1996--2000. In the present 
work we have used only the five spectra secured in 2006--2007, as the orbital parameters gradually change 
due to apsidal motion.

\subsection{La Palma spectra}

15 spectra of V380\,Cyg were obtained in November 2006 with the Nordic Optical Telescope (NOT) and FIES 
\'echelle spectrograph (Frandsen \& Lindberg 1999) at La Palma. We used the medium-resolution fibre which 
yielded a resolving power of $R = 47\,000$ and a fixed wavelength coverage of 3640--7360\,\AA. Wavelength 
calibration was performed using thorium-argon exposures, and flat-fields were obtained using a halogen lamp. 
The observing conditions were generally reasonable (for winter  in La Palma) but several exposures suffered 
from thin cloud coverage or poor seeing.

\subsection{Data reduction}

The \'echelle spectra (from CAHA and NOT) were bias subtracted, flat-fielded and extracted with the 
{\sc iraf}\footnote{{\sc iraf} is distributed by the National Optical Astronomy Observatory, which are 
operated by the Association of the Universities for Research in Astronomy, Inc., under cooperative agreement
 with the NSF.} \'echelle package routines. Normalisation and merging of the orders was performed with 
great care, using programs developed by ourselves, to ensure that these steps did not causes any systematic
 errors in the resulting spectra.

The Ond\v{r}ejov spectra were reduced by the program {\sc spefo} (Horn et al.\ 1996, \v{S}koda 1996). 
Initial reduction of the DAO spectra (bias subtraction, flatfielding, and spectrum extraction) were carried
 out in {\sc iraf}. The wavelength calibrations were based on thorium-argon lamp spectra, and were 
performed using {\sc spefo}.

%%%%%%%%%%%%%%%%%%%%%%%%%%%%%%%%%%%%%%%%%%%%%%%%%%%%%%%%%%%%%%%%%%%%%%%%%%%%%%%%%%%%%%%%%%%%%%%%%%%%%%%%%%%%%%%%%%

\section{Method}                                                                                \label{sec:method}

Our analysis follows the methods introduced by Hensberge, Pavlovski \& Verschueren (2000) and PH05 in 
their studies of the eclipsing and double-lined spectroscopic binary V578\,Mon, and elaborated in Paper\,I 
(Pavlovski \& Southworth 2009). The core of this approach is reconstruction of individual stellar spectra 
from the observed composite spectra using {\sc spd}, which allows a detailed abundance analysis using the 
same tools as for single stars (c.f.\ Pavlovski 2004, Hensberge \& Pavlovski 2007). {\sc spd} also gives 
the velocity amplitudes of the two stars which, in combination with modelling of V380\,Cyg's light curves, 
yields the physical properties of the two stars including precise and accurate surface gravity measurements 
which are vital for the abundance analyses.

The method of {\sc spd} returns the individual spectra of the two components of a binary star, but their 
normalisation is arbitrary. This is because the continuum light ratio of the two stars is {\em not} directly 
measurable from composite spectra, short of ensuring individual spectral lines do not dip below zero flux. 
Whilst relative line strengths in a spectrum are reliable, their absolute scaling must be recovered from 
elsewhere. Here we have renormalised our disentangled spectra using the stellar light ratios found in the 
light curve analysis  (Sec.\,\ref{sec:lc}), following the procedure described in detail by PH05, and taking 
into account the line-blocking inherent to the Fourier disentangling method.

We have estimated chemical abundances by fitting the renormalised disentangled spectra with synthetic 
spectra. Non-LTE line formation and spectrum synthesis computations were performed using {\sc detail} and 
{\sc surface} (Giddings 1981, Butler 1984) and the model atoms as listed in Paper\,I. Model atmospheres 
were calculated with {\sc atlas9} (Kurucz 1979). Justification for this hybrid approach can be found in 
Nieva \& Przybilla (2007).

{\sc spd} can be quite sensitive to the precise phase distribution and continuum normalisation of the 
observed composite spectra. Hynes \& Maxted (1998) have shown that this is not a problem for radial 
velocity studies, but Hensberge, Iliji\'{c} \& Torres (2008) have demonstrated that it is important for 
correct reproduction of the overall shape of disentangled spectra. A lack of significant time-variability 
in the relative light contributions of the two star can produce spurious patterns in separated spectra, 
which in practise appear as typically low-frequency continuum variations. Hensberge et al.\ (2008) 
examined this in detail and found that the phenomenon arises due to degeneracy of the {\sc spd} 
equations and/or bias progression from the observed spectra. Careful planning of an observing run 
and the subsequent data reduction and analysis can mostly overcome these difficulties, in particular 
by observing the target during eclipses. In the case of V380\,Cyg we were hampered by the relative 
faintness of the secondary star, but were able to secure spectra with a very good distribution in 
orbital phase. Combined with a careful data reduction (particularly concerning continuum normalisation 
and \'echelle order merging), this approach has allowed us to obtain high-quality disentangled spectra 
which can be used for abundance analysis.

%%%%%%%%%%%%%%%%%%%%%%%%%%%%%%%%%%%%%%%%%%%%%%%%%%%%%%%%%%%%%%%%%%%%%%%%%%%%%%%%%%%%%%%%%%%%%%%%%%%%%%%%%%%%%%%%%%

\section{Spectroscopic orbits through spectral disentangling}                                   \label{sec:orbits}

In the method of {\sc spd} as introduced by Simon \& Sturm (1994), one solves both for the individual 
spectra of the components, and the optimal set of  orbital parameters. Here we have performed a {\spd}
 analysis in Fourier space  as formulated by Hadrava (1995), which has one big advantage. The system 
of equations consists of $N_{\rm obs} \times N_{\rm pix}$ coupled equations linear in a little more 
than $2N_{\rm pix}$, where $N_{\rm obs}$ is the number of observed spectra, and $N_{\rm pix}$ is the 
number of datapoints per each observed spectrum. Using discrete Fourier transforms this large system 
of equations can be coupled into $(N_{\rm pix}/2) + 1$ systems of $N_{\rm obs}$ complex equations 
involving only two unknowns in the case of binary systems, and the required computation time is 
significantly smaller. In this work we have performed {\sc spd} on a large spectral range about 
3000\,\AA\ wide, for which the speed advantage of Fourier disentangling is very important. The 
disadvantage is that weights can be assigned only to whole spectra, and not to individual pixels. 
We weighted the spectra according to S/N. We use the 
{\sc fdbinary}\footnote{\tt http://sail.zpf.fer.hr/fdbinary/} code described 
by Iliji\'{c} et al.\ (2004).

{\spd} was performed in spectral regions centred on the prominent \ion{He}{I} lines. The contribution 
of star B to the total system light is rather small ($L_{\rm A}/L_{\rm B} \sim 15.5$ in $V$), and its 
lines are difficult to see. Fortunately, star B has a similar \Teff\ to star A 
(Sec.\,\ref{sec:analysis}), and so is best seen in the \ion{He}{I} lines. Star B's lines also 
emerge in other regions of its disentangled spectrum, but are very weak so are of limited use 
for determining orbital elements. We therefore concentrated only on the stronger helium lines. 
After some trial calculations, we selected six spectral regions centred on the lines at 
$\lambda\lambda$ 4026, 4388, 4471 (including \ion{Mg}{II} $\lambda$4481), 4712, 4912, and 6670 {\AA} 
for measuring the orbital elements. We also grouped the spectra into appropriate time intervals to 
account for the change in $\omega$ due to apsidal motion (amounting to about 6$^\circ$ between our 
first and last spectrum). Table\,\ref{tab:orbit} lists the final orbital elements, which are the mean 
and rms of the values derived from all spectral regions.

\begin{table} \centering \caption{\label{tab:orbit} Parameters
of the spectroscopic orbits for V380\,Cyg derived in this work.
Throughout the {\spd} analysis the period was kept fixed to the
value derived by G2000.}
\begin{tabular}{llll} \hline
Orbital period (d)                          & 12.425719 (fixed)  \\
Time of periastron passage (HJD)            & $2454615.18 \pm 0.14$ \\
Velocity amplitude $K_{\rm A}$ (\kms)       & $95.1 \pm 0.3$     \\
Velocity amplitude $K_{\rm B}$ (\kms)       & $160.5 \pm 1.2$    \\
Mass ratio $q$                              & $0.592 \pm 0.005$  \\
Orbital eccentricity $e$                    & $0.206 \pm 0.008$  \\
Longitude of periastron $\omega$ ($^\circ$) & $134.2 \pm 1.1$    \\
\hline \\
\end{tabular}
\end{table}

\begin{figure} \centering \includegraphics[width=85mm]{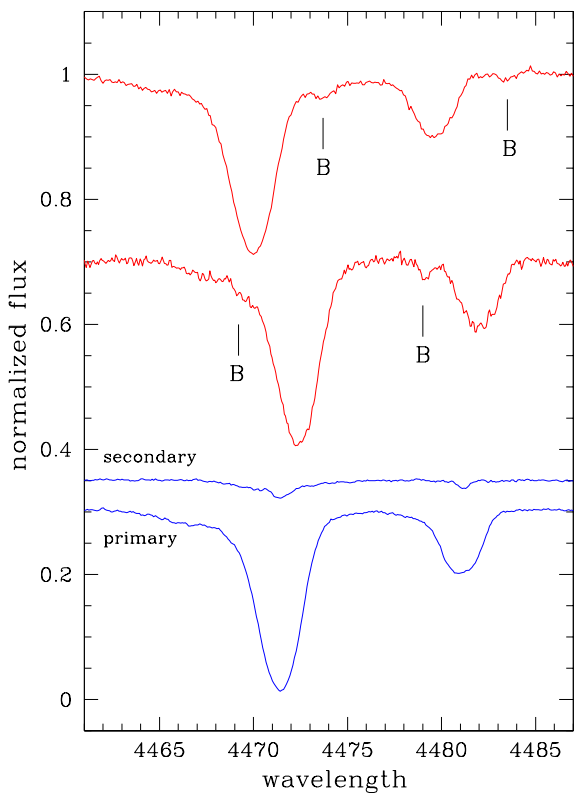} \\
\caption{\label{fig:plotspe} Observed spectra around the \ion{He}{I}
$\lambda$4471 and \ion{Mg}{II} $\lambda$4481 lines. The upper two
spectra were obtained near quadrature, and weak lines of the
secondary are visible and indicated with a `B'. The lower two
spectra are disentangled spectra of the components. The faintness
of star B compared to star A is obvious.} \end{figure}

How does our solution compare to previous results? Batten (1962) made a comprehensive study of all 
available photographic spectra, and measured the lines of star B for the first time. He found 
$K_{\rm A} = 93.4 \pm 2.0$\kms\ and $K_{\rm B} = 160.9 \pm 2.9$ \kms, which are very close to our 
values. Hill \& Batten (1984) studied existing and new photographic spectra, and found a discrepancy
 between the systemic velocities of the two stars. Their $K_{\rm B}$ is about 7\kms\ larger than 
Batten's, which increases the mass of star A from 12.5 to 13.7 \Msun. Lyubimkov et al.\ (1996) 
collected all available RV measurements for V380\,Cyg, including new CCD observation. They found 
a much smaller $K_{\rm B}$ of 155.3\kms\ and a slightly larger $K_{\rm A}$ of 93.95\kms, but gave 
no uncertainties. The most comprehensive investigation to date is that of Popper \& Guinan (1998), 
based on high-resolution and high-S/N \'echelle spectra. They adopted the average of their own and 
Lyubimkov's solutions, arriving at $K_{\rm A} = 95.6 \pm 0.5$\kms\ and $K_{\rm B} = 151 \pm 3$\kms. 
Our own results compare well for $K_{\rm A}$, but a discrepancy of 10\kms\ exists for $K_{\rm B}$.

The main source of error in RV measurement by cross-correlation is the selection of the template 
spectrum. {\sc spd} bypasses this problem entirely, by directly seeking the best overall solution 
for both the spectra and the orbital elements. It therefore does not require guidance from a template 
spectrum, or pass through an intermediate step of obtaining RV measurements from each observed spectrum. 
An additional advantage of this approach is that it is not biased by line blending. Unfortunately, 
a dedicated study of error propagation in {\spd} is still lacking, and the best we can do is split 
our spectra into different sets and different wavelength regions to obtain several independent 
measurements of the orbital parameters (Table\,\ref{tab:orbit}). Recently, Southworth \& Clausen 
(2007) have performed an extensive analysis of orbits  derived by double Gaussian fitting, 
cross-correlation, and {\sc spd}. The {\sc spd} orbits were the most internally consistent, 
followed by double Gaussian fitting. The cross-correlation results were affected by line blending, 
even after attempts were made to correct for this. Southworth \& Clausen (2007) noticed that their 
{\sc spd} solutions had several local minima around the best fit, and used a grid search to deal 
with this.

We therefore ascribe the 10\kms\ discrepancy in $K_{\rm B}$ to line blending in the cross-correlation
 solutions. Our $K_{\rm B}$ was found by {\sc spd}, which does not suffer from line blending or 
template errors, and should be preferred. Finally, as line blending is not a problem for {\sc spd}, 
we could use all of our spectra rather than just those where the lines of the two stars are resolved. 
Disentangled profiles of the \ion{He}{I} $\lambda$4471  and \ion{Mg}{II} $\lambda$4481 lines are 
shown in Fig.\,\ref{fig:plotspe}.

%%%%%%%%%%%%%%%%%%%%%%%%%%%%%%%%%%%%%%%%%%%%%%%%%%%%%%%%%%%%%%%%%%%%%%%%%%%%%%%%%%%%%%%%%%%%%%%%%%%%%%%%%%%%%%%%%%

\section{Light curve analysis}                                                                      \label{sec:lc}

Driven by suspicions that the secondary star of V380\,Cyg might be hotter than the primary star 
(see below), we have revisited the light curves presented by G2000. These data were obtained using
 the Automated Photometric Telescopes at Mt.\ Hopkins, Arizona, and total 870 observations in the 
$UBV$ bandpasses.

The light curves show substantial ellipsiodal modulation and reflection effect, so we have used 
the Wilson-Devinney ({\sc wd}) code (Wilson \& Devinney 1971; Wilson 1979, 1993), which implements 
Roche geometry and a detailed treatment of reflection and other physical phenomena, version of 
2004/02/06. We have modified {\sc wd} to automatically converge to the best solution using either
 a damped version of the standard differential corrections procedure ({\sc dc}) or the robust 
downhill simplex algorithm {\sc amoeba} (Press et al.\ 1992). For our final solutions we used 
{\sc dc} and iterated until all parameter corrections were less than half of their formal errors.

The full set of fixed and control parameters are given in Table\,\ref{tab:wdfix}: for our modelling 
we adopted the orbital ephemeris from G2000, bolometric albedos and gravity brightening exponents
 appropriate for radiative atmospheres (Claret 1998, Claret 2001) and pseudosynchronous rotation 
for both stars. Using the detailed reflection technique of Wilson (1990) did not substantially 
improve the fit and much increased the calculation time, so was not used.

The treatment of limb darkening can be important, and there is evidence that theoretically 
predicted coefficients are imperfect (Southworth 2008; Southworth et al.\ 2009). We therefore 
included limb darkening in three different ways: using coefficients obtained by bilinear 
interpolation in the tables of Van Hamme (1993) or Claret (2000), or including the coefficients
 as fitted parameters. The solutions show a negligible dependence on the way limb darkening 
is accounted for, probably because the eclipses are quite shallow compared to the observational 
scatter. For our final solutions we used the linear limb darkening law and fixed the coefficients 
at the Van Hamme (1993) values (Table\,\ref{tab:wdfix}).

\begin{table} \centering \caption{\label{tab:wdfix} Summary of the
fixed and control parameters for our final {\sc wd} solutions of the
G2000 light curves of V380\,Cyg. For further details please see the
{\sc wd} user guide (Wilson \& Van Hamme 2004).}
\begin{tabular}{llcc} \hline
Parameter                 & {\sc wd} name      & Star A & Star B   \\
\hline
Orbital period (d)        & {\sc period}       & \mc{12.425719}    \\
Reference time (HJD)      & {\sc hjd0}         & \mc{2441256.544}  \\
Mass ratio                & {\sc rm}           & \mc{0.5916}       \\
Stellar \Teff s (K)       & {\sc tavh, tavc}   & 21\,500 & 22\,000 \\
Rotation rates            & {\sc f1, f2}       & 1.0     & 1.0     \\
Stellar albedos           & {\sc alb1, alb2}   & 1.0     & 1.0     \\
Gravity darkening         & {\sc gr1, gr2}     & 1.0     & 1.0     \\
Numerical accuracy        & {\sc n1, n2}       & 60      & 30      \\
Bolometric LD coefficient & {\sc xbol1, xbol2} & 0.6475  & 0.6854  \\
$U$ LD coefficient        & {\sc x1, y1}       & 0.3582  & 0.3128  \\
$B$ LD coefficient        & {\sc x1, y1}       & 0.3534  & 0.3026  \\
$V$ LD coefficient        & {\sc x1, y1}       & 0.3043  & 0.2622  \\
\hline \end{tabular} \end{table}

\subsection{The effective temperatures}

\begin{figure} \includegraphics[width=0.48\textwidth,angle=0]{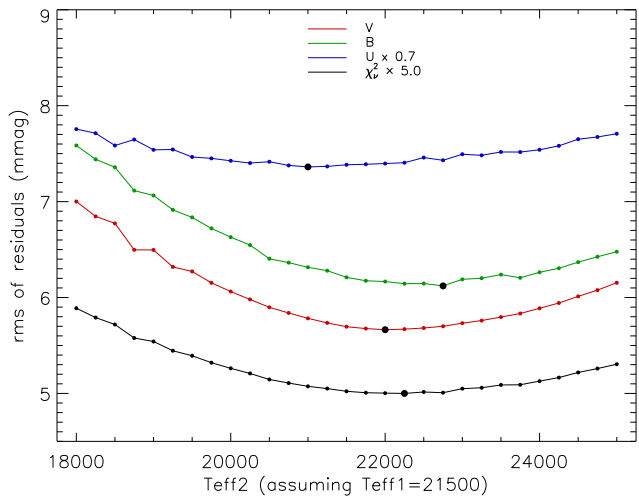}
\caption{\label{fig:teffrms} Quality of the solutions for the $UBV$ light
curves individually and in combination, for ${\Teff}_{\rm A} = 21\,500$\,K
and ${\Teff}_{\rm B}$ fixed at a range of values. For the individual light curves
the solution quality is given by the rms of the residuals, and for the combined
solution the reduced $\chi^2$ assuming observational errors equal to the rms value
for each light curve. The points at which a fit was evaluated or the best fit was
found are given by small and large filled circles, respectively.} \end{figure}

The light curves of dEBs contain almost no information about the absolute \Teff s of the 
component  stars: only the {\em ratio} of the \Teff s is directly derivable from normal 
light curves. One of our goals was to investigate the quality of solutions where star B 
was hotter than star A. For one set of solutions we therefore  fixed ${\Teff}_{\rm A} = 
21\,500$\,K and then fixed ${\Teff}_{\rm B}$ to values between $18\,000$ and $25\,000$\,K 
in steps of $250$\,K. The best fit was found for each ${\Teff}_{\rm B}$, and for the three 
light curves both individually and in combination. Fig.\,\ref{fig:teffrms} is a plot of the 
quality of the fits for these solutions, and shows a preference for solutions with 
${\Teff}_{\rm B} >  {\Teff}_{\rm A}$.

\subsection{Final light curve solutions}

\begin{table*} \centering \caption{\label{tab:lcfit} Results of the {\sc wd}
code modelling process of the G2000 light curves of V380\,Cyg. For the final
column the uncertainties come from the scatter of the individual solutions
for the $U$, $B$ and $V$ data. For the other columns the uncertainties are
formal errors calculated by the {\sc wddc} code.}
\begin{tabular}{ll r@{\,$\pm$\,}l r@{\,$\pm$\,}l r@{\,$\pm$\,}l r@{\,$\pm$\,}l r@{\,$\pm$\,}l} \hline
Parameter                   &{\sc wd} name&    \mc{$U$}      &    \mc{$B$}      &    \mc{$V$}      & \mc{Combined}    &  \mc{Adopted}      \\
\hline
Star A potential            & {\sc phsv}  &   4.698 & 0.073  &   4.541 & 0.030  &   4.587 & 0.039  &   4.609 & 0.029  &   4.609  & 0.046  \\
Star B potential            & {\sc pcsv}  &  10.75  & 0.25   &  10.51  & 0.14   &  10.26  & 0.14   &  10.51  & 0.12   &  10.51   & 0.17   \\
Orbital inclination (\degr) & {\sc xincl} &  80.76  & 0.58   &  80.63  & 0.17   &  81.25  & 0.31   &  81.01  & 0.21   &  81.01   & 0.27   \\
Orbital eccentricity        & {\sc e}     &   0.2207& 0.0042 &   0.2023& 0.0021 &   0.1831& 0.0017 &   0.1979& 0.0017 &   0.198  & 0.012  \\
Periastron longitude (\degr)& {\sc perr0} & 136.5   & 2.7    & 140.3   & 1.7    & 150.1   & 2.1    & 143.2   & 1.5    & 143.2    & 5.3    \\
Phase shift                 & {\sc pshift}&$-$0.0459& 0.0009 &$-$0.0460& 0.0004 &$-$0.0480& 0.0004 &$-$0.0470& 0.0004 &     \mc{ }        \\
$U$-band light from star A  & {\sc hlum}  &  12.008 & 0.046  &     \mc{ }       &     \mc{ }       &  12.055 & 0.027  &     \mc{12.055}   \\
$U$-band light from star B  & {\sc clum}  &   0.734 & 0.045  &     \mc{ }       &     \mc{ }       &   0.697 & 0.026  &     \mc{ 0.697}   \\
$B$-band light from star A  & {\sc hlum}  &    \mc{ }        &  9.443  & 0.023  &     \mc{ }       &   9.447 & 0.025  &     \mc{ 9.447}   \\
$B$-band light from star B  & {\sc clum}  &    \mc{ }        &  0.618  & 0.022  &     \mc{ }       &   0.611 & 0.024  &     \mc{ 0.611}   \\
$V$-band light from star A  & {\sc hlum}  &    \mc{ }        &     \mc{ }       &   9.823  & 0.024 &   9.805 & 0.025  &     \mc{ 9.805}   \\
$V$-band light from star B  & {\sc clum}  &    \mc{ }        &     \mc{ }       &   0.619  & 0.023 &   0.635 & 0.025  &     \mc{ 0.635}   \\
Fractional radius of star A      &        &    \mc{0.2563}   &     \mc{0.2660}  &     \mc{0.2615}  &     \mc{0.2609}  &   0.2609 & 0.0040 \\
Fractional radius of star B      &        &    \mc{0.06391}  &     \mc{0.06538} &     \mc{0.06701} &     \mc{0.06534} &   0.0653 & 0.0013 \\
\multicolumn{2}{l}{$U$-band rms of residuals (mmag)}         &    \mc{10.386}   &     \mc{ }       &     \mc{ }       &     \mc{10.524}  &     \mc{ }        \\
\multicolumn{2}{l}{$B$-band rms of residuals (mmag)}         &    \mc{ }        &     \mc{5.964}   &     \mc{ }       &     \mc{6.130}   &     \mc{ }        \\
\multicolumn{2}{l}{$V$-band rms of residuals (mmag)}         &    \mc{ }        &     \mc{ }       &     \mc{5.782}   &     \mc{5.635}   &     \mc{ }        \\
\hline \end{tabular} \end{table*}

\begin{figure} \includegraphics[width=0.48\textwidth,angle=0]{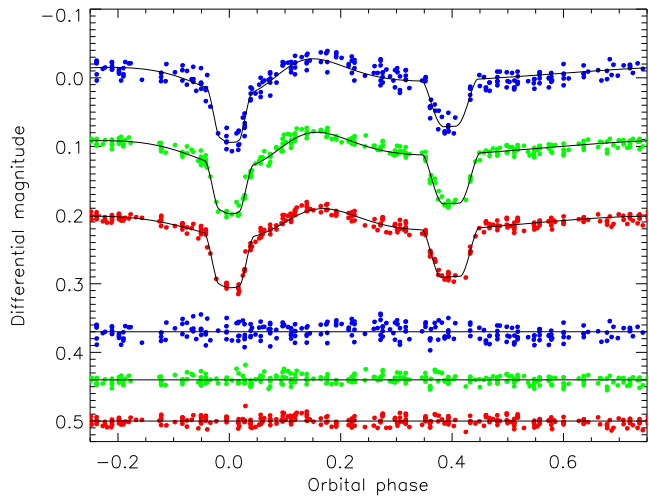}
\caption{\label{fig:lc} Comparison between the best fits found with the {\sc wd}
code and the light curves presented by G2000. The residuals are shown at the
base of the plot with offsets from zero. In each case the top (blue) data are
$U$, the middle (green) are $B$ and the lower (red) are $V$.} \end{figure}

We have obtained final solutions for the $UBV$ light curves both individually and in combination, 
using our damped version of {\sc dc} to optimise the fits. {\sc wd} allows the use of predictions 
from Kurucz (1993) model atmospheres to link the flux ratios in different passbands, which means 
the results have a dependence on theoretical models. We prefer to avoid this option ({\sc mode}$=$0 
and {\sc ipb}$=$1; Wilson \& Van Hamme 2004), particularly as the available data are multi-band and 
include the very blue $U$ passband. We therefore fit for the light contribution of each star in each
 passband independently, after assuming reasonable \Teff s with which the limb darkening coefficients 
are derived.

We fixed ${\Teff}_{\rm A} = 21\,500$\,K and ${\Teff}_{\rm B} = 22\,000$\,K, and then fitted for the 
potentials of the two stars, the orbital inclination, eccentricity and periastron longitude, a phase 
shift, and the light contributions of the two stars in each passband. Additional light from a third 
star was not considered (see G2000).

The three light curves were fitted together to give the final parameter values, and then individually
 to check for consistency. In Table\,\ref{tab:lcfit} we report the fitted parameters and their 
formal errors (calculated by {\sc dc} from the covariance matrix), and in Fig.\,\ref{fig:lc} we 
show the fits to the data. The volume-equivalent fractional radii of the two stars were obtained 
using the {\sc wd} {\sc lc} program.

We caution that the formal errors can be optimistic when there are strong correlations between
 parameters, so should be considered with care. Taking the formal errors at face value indicates 
poor agreement between light curves, with reduced $\chi^2$ values of 4.2 and 3.4 for the stellar 
potentials, 3.4 for the inclination and a huge 96.9 and 19.7 for $e$ and $\omega$. This highlights 
the limitations of formal errors, as $e$ and $\omega$ are strongly correlated in situations such 
as this one (e.g.\ Southworth et al.\ 2004, 2007). For our final light curve parameters we 
instead adopt the values from the combined fit to the three light curves and the uncertainties 
from the scatter between the separate solutions for these light curves.

%%%%%%%%%%%%%%%%%%%%%%%%%%%%%%%%%%%%%%%%%%%%%%%%%%%%%%%%%%%%%%%%%%%%%%%%%%%%%%%%%%%%%%%%%%%%%%%%%%%%%%%%%%%%%%%%%%

\section{Spectral analysis of both components}                                                \label{sec:analysis}

A model atmosphere is defined to first order by its \Teff\ and \logg, and when fitting 
observational data these two parameters can be quite correlated. This can be a major limition 
for abundance analyses of single stars, but in the case of dEBs it is possible to accurately 
measure their \logg s from light and velocity curve analysis. In the case of V380\,Cyg we have
 measured ${\logg}_{\rm A} = 3.136 \pm 0.014$ and ${\logg}_{\rm B} = 4.112 \pm 0.017$ 
(Sec.\,\ref{sec:absdim}).

\subsection{Effective temperature determination}                                                  \label{sec:teff}

The \Teff s of B stars are best determined from the silicon ionisation balance (Becker 
\& Butler 1990). Lines of \ion{Si}{II} and \ion{Si}{III} are present in the disentangled 
spectrum of V380\,Cyg\,A -- the lack of \ion{Si}{IV} lines indicates that ${\Teff}_{\rm A} 
< 24\,000$\,K. Grids of synthetic spectra were calculated for $\Teff = 19000$--$24000$\,K 
and for $\logg = 3.136$. Equivalent widths (EWs) were measured in {\sc iraf} for the 
\ion{Si}{II} ($\lambda\lambda$ 4128, 4130, 5031) and \ion{Si}{III} ($\lambda\lambda$ 
4552, 4567, 5039) lines, and calibration curves produced for \ion{Si}{II}/\ion{Si}{III}
 EW ratios. EW measurements were made for star A and a mean value of ${\Teff}_{\rm A} = 21\,750 
\pm 220$\,K was found from six EW ratios. Absolute EWs can only be measured from disentangled 
spectra once they have been carefully renormalised, but the {\em ratios} of EWs are reliable.

\begin{table*} \centering \caption{\label{tab:genfit}
Results of optimised genetic algorithm fitting of the hydrogen Balmer lines of star A.
In the second column the parameter which was optimised is indicated; parameters were
otherwise fixed to $\log g = 3.131$ (and $\ell_{\rm A} = 1$ for the renormalised
spectrum). The light curve analysis gave $\ell_{\rm A} = 0.9394$. For comparison,
analysis of the silicon ionisation balance gives ${\Teff}_{\rm A} = 21\,750 \pm
220$\,K. In the last column we give the difference between the value obtained in
optimised fitting to the value adopted in this work. $\chi^2$ is also given.}
\begin{tabular}{clllllr} \hline
Run & Parameter & H$\beta$        & H$\gamma$        & H$\delta$        &  Mean  &  Difference  \\ \hline
1  & \Teff     & 21\,960$\pm$85   & 22\,270$\pm$70   & 21\,860$\pm$55   & 22\,030$\pm$210   & 280    \\
   & ${\chi}^2$& 0.0037144        & 0.0036232        & 0.0038175        & -                 &        \\[2pt]
2  & \Teff     & 21\,850$\pm$55   & 21\,690$\pm$31   & 21\,800$\pm$45   & 21\,780$\pm$80    & 30     \\
   & $\log g$  & 3.121$\pm$0.034  & 3.125$\pm$0.031  & 3.123$\pm$0.030  & 3.123$\pm$0.030   & 0.008  \\
   & ${\chi}^2$& 0.0025443        & 0.0023554        & 0.0027782        & -                 &        \\[2pt]
3  & \Teff     & 21\,830$\pm$70   & 21\,620$\pm$80   & 21\,970$\pm$75   & 21\,764$\pm$43    & 14     \\
   & $\ell $   & 0.9541$\pm$0.0021& 0.9471$\pm$0.0034& 0.9487$\pm$0.0029& 0.9512$\pm$0.0015 & 0.0118 \\
   & ${\chi}^2$& 0.0019254        & 0.0018643        & 0.0020095        & -                 &        \\[2pt]
4  & \Teff     & 21\,920$\pm$95   & 21\,890$\pm$90   & 21\,845$\pm$95   & 21\,885$\pm$54    & 135    \\
   & $\log g$  & 3.122$\pm$0.005  & 3.129$\pm$0.007  & 3.139$\pm$0.004  & 3.136$\pm$0.003   & 0.005  \\
   & $\ell $   & 0.9677$\pm$0.0026& 0.9501$\pm$0.0041& 0.9675$\pm$0.0031& 0.9643$\pm$0.0018 & 0.0249 \\
   & ${\chi}^2$& 0.0015442        & 0.0015135        & 0.0016024        & -                 &        \\
\hline \end{tabular} \\
\end{table*}

\begin{figure*} \centering
\includegraphics[width=45mm]{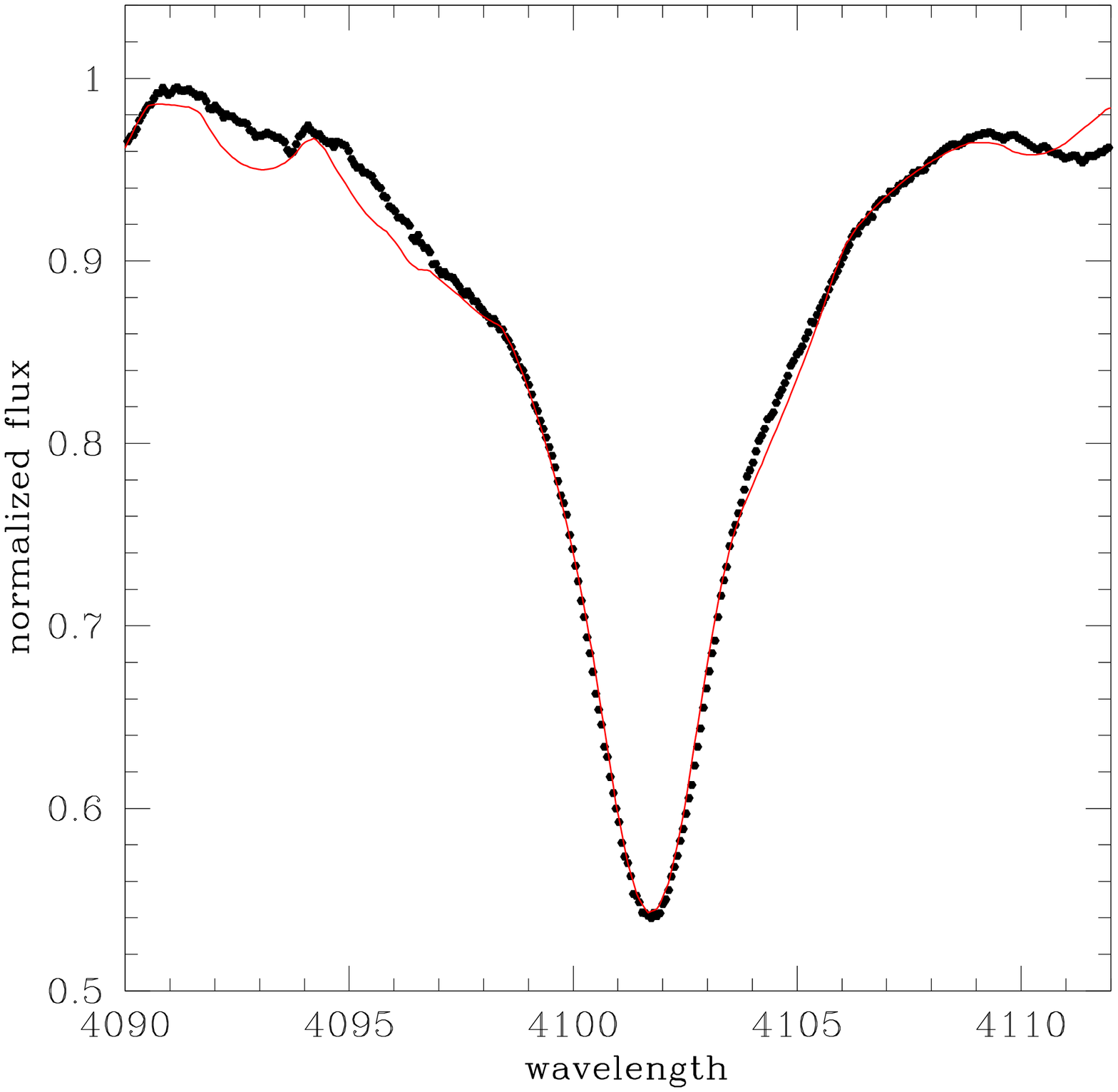} \
\includegraphics[width=45mm]{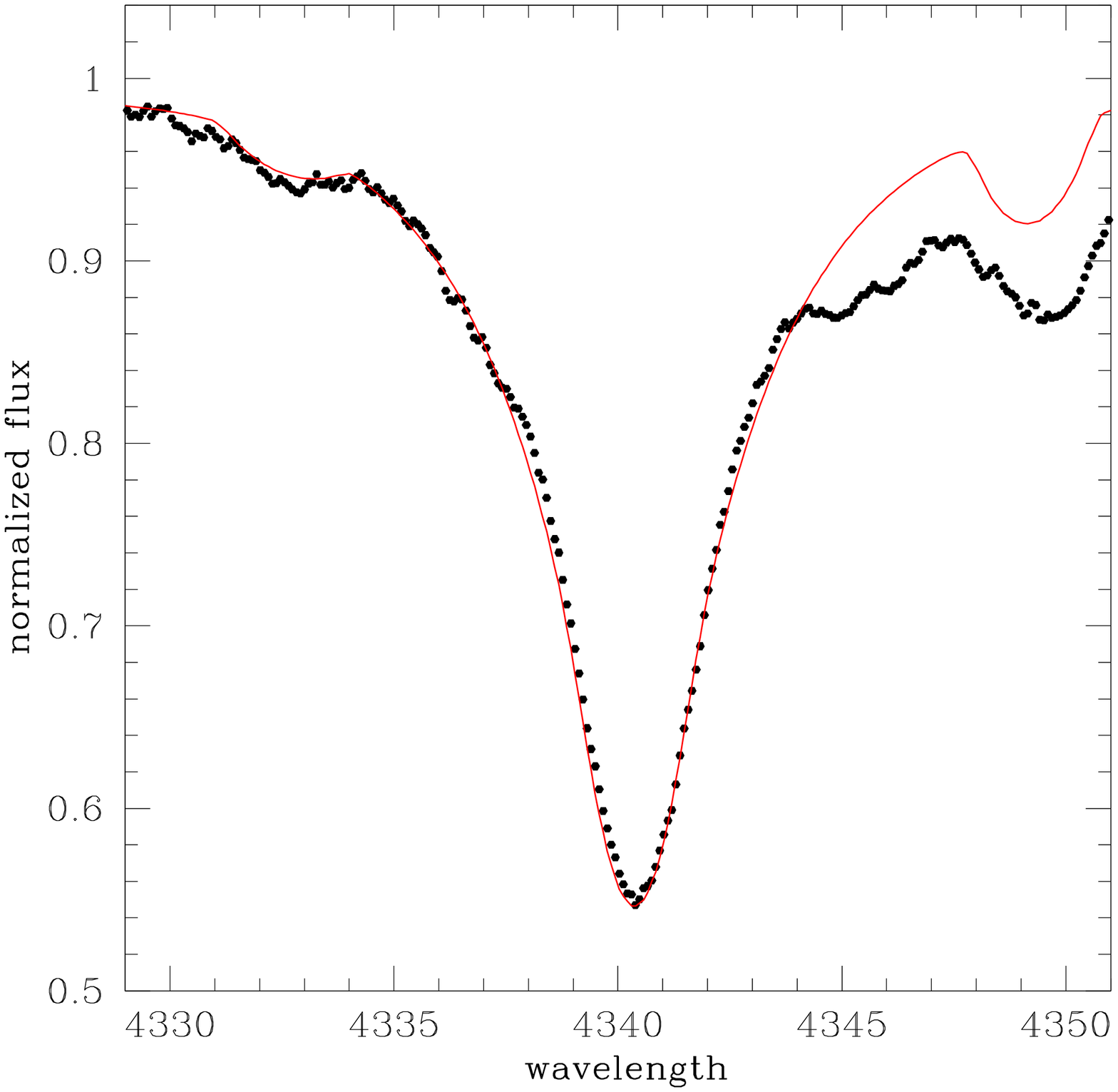} \
\includegraphics[width=45mm]{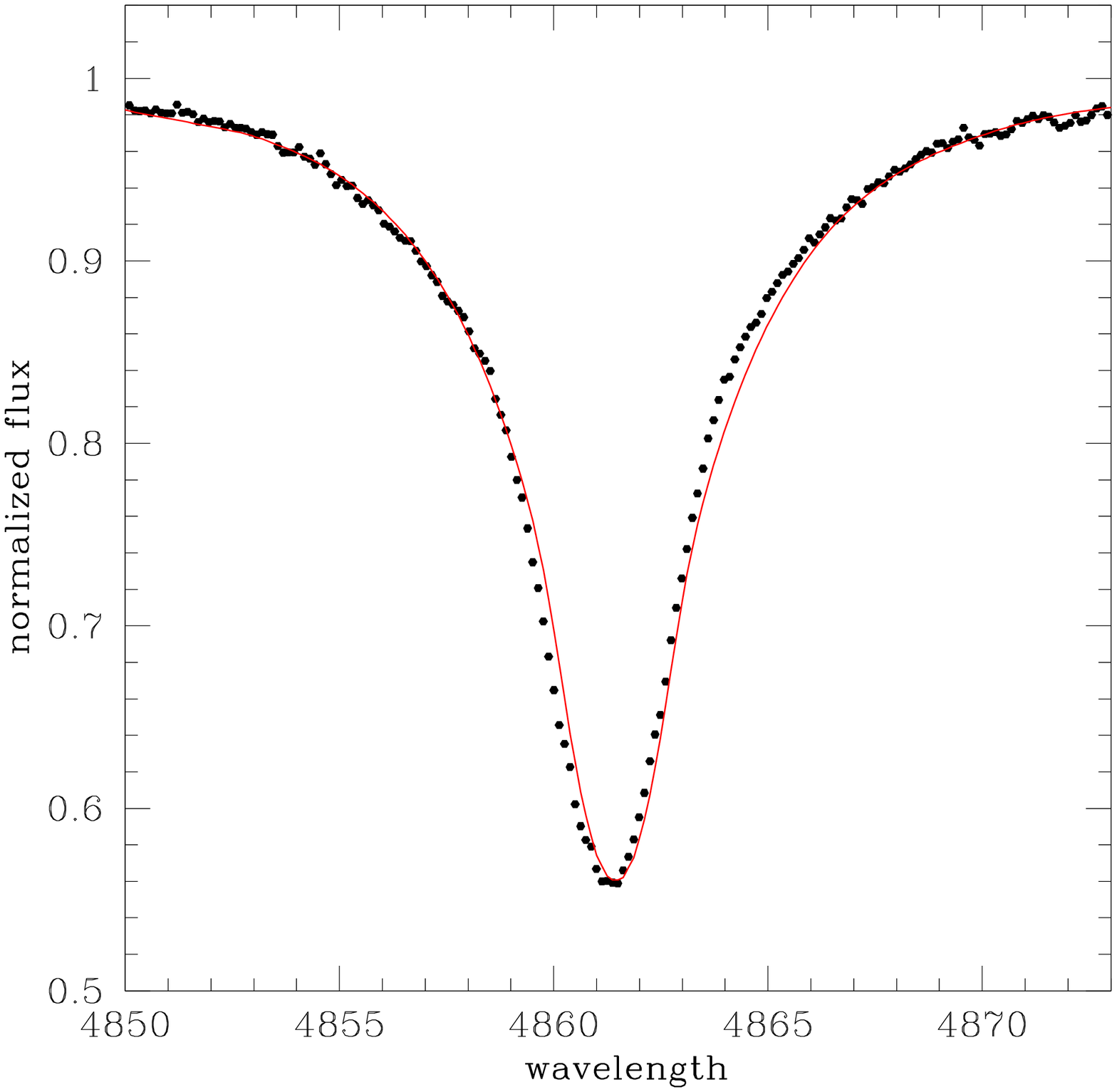}  \\
\caption{\label{fig:fitbalmer} Comparison between the disentangled
spectra (points)  best-fitting theoretical spectra (solid lines) for
H$\delta$ (left), H$\gamma$ (middle) and H$\beta$ (right). } \end{figure*}

A second way of measuring \Teff s of B stars is to study the shape of the hydrogen Balmer lines. 
The usual degeneracy between \Teff\ and \logg\ is not a problem here, as the \logg\ values of the
 stars are known, but this method does require a careful renormalisation of the disentangled 
spectra. We used light factors derived from the light curve analysis (Sec.\,\ref{sec:lc}), whose
 variation with wavelength is small in the region covered by H$\delta$, H$\gamma$ and H$\beta$. 
We avoid H$\alpha$ since it is formed high in the stellar atmosphere and is not a good \Teff\ 
indicator. The fitting was done with a routine for multi-parameter optimisation, which uses 
a genetic algorithm to minimise $\chi^2$ (Tamajo et al.\ 2009). The projected rotational velocity 
was fixed to $v\sin i = 98$ \kms\ (Sec.\,\ref{sec:micro}). Runs were performed in which just 
\Teff\ (run\,1) or both \Teff\ and \logg\ (run\,2) were optimised. The results are tabulated 
in Table\,\ref{tab:genfit} and compared to the observed spectra in Fig.\,\ref{fig:fitbalmer}.

We also tested the inclusion of the light factor as a fitted parameter (using the separated
but not renormalised spectrum) with encouraging results (run\,3 and run\,4 in Table\,\ref{tab:genfit}).
This suggests that it is possible to reliably estimate \Teff s using disentangled spectra for SB2 systems
which are {\rm not} eclipsing (so do not have well-determined light contributions from their component
stars through eclipse analysis), which is a very interesting possibility. The results also corroborate
the light contributions found in Sec.\,\ref{sec:lc}.

The two methods agree reasonably well on the value of ${\Teff}_{\rm A}$, although the difference 
is slightly larger than 1$\sigma$ for run\,1. Possible sources of systematic error include the 
placement of the continuum for the EW measurements (particularly for the weak \ion{Si}{II} 
$\lambda$4218 and $\lambda$4231 lines), blending between \ion{Si}{II} $\lambda$4231 and 
\ion{O}{II} $\lambda$4233, and undulations in the continuum in the region of the (wide) 
Balmer lines. 
Systematic errors could also arise from the adopted atomic data (both 
in the statistical equilibrium calculations and in the spectral synthesis) and in the LTE 
assumptions of spatial homogenity, plane parallel geometry, radiative and hydrostatic 
equilibrium, etc. However, a study of B-type supergiants by Trundle et al.\ (2004), using 
sophisticated non-LTE analyses, yielded \Teff\ measurements in good agreement with the 
calibration of Dufton et al.\ (2000), which uses assumptions similar to those listed above. 
Despite this, and the generally good ageement we find between the \Teff s obtained in 
different ways, we have to be cautious as it is difficult to quantify the size of the 
systematic errors. We note that the high precison of our \Teff\ values comes partly from 
having accurate and precise \logg\ values for the two stars. As the final we adopted
${\Teff}_{\rm A} =  21\,750 \pm 280$ K where error accounts for the determination
of \Teff\ from hydrogen line profiles, also.

\subsection{Microturbulence and projected rotational velocity}     \label{sec:micro}

We have estimated the microturbulence velocity, \micro, by requiring that the measured abundance 
of a chemical element does not depend on the strength of its lines: the slope of a plot of EW 
versus abundance should be zero. The \ion{O}{II} lines are the most numerous in the spectrum 
of star A, allowing us to use 32 lines for measuring \micro. We find $\micro = 14 \pm 1$\kms, 
which is relatively high but not unusual for evolved stars with $\log g \sim 3$ (Hunter et al.\ 2007, 
Morel et al.\ 2006). It is encouraging that G2000 derived $\micro = 12\pm 1$\kms\ using a 
completely different approach.

The difficulty of determining \micro\ has been discussed recently by Hunter et al.\ (2008). 
For B stars, the rich spectrum of the \ion{O}{II} ion is often used for this purpose. 
Uncertainties can arise between different multiplets due to errors in the adopted atomic 
data or in the strength of non-LTE effects, so using a single \ion{O}{II} multiplet is best. 
Dufton et al.\ (2005) preferred the \ion{Si}{III} $\lambda$4560 triplet, and this approach 
was adopted by Hunter et al.\ (2008) in a large VLT-FLAMES survey of B stars. We have also 
tried this approach, and found almost the same value, $\micro = 13 \pm 1$\kms, as above. 
The use of an alternative oxygen multiplet at 4072, 4076 and 4079 {\AA} was less successful 
as the $\lambda$4079 line is weak and so made very shallow by rotational broadening. Therefore,
 an error in \micro of $\pm 2$ \kms\ would be more realistic.

The projected rotational velocities ($v \sin i$) of the components of V380\,Cyg were derived 
from the widths of several clean spectral features: \ion{C}{II} $\lambda$4267, \ion{Si}{III} 
$\lambda$4552, \ion{Si}{III} $\lambda$4567, \ion{O}{II} $\lambda$4591, \ion{O}{II} 
$\lambda$4596, and \ion{O}{II} $\lambda$4662; and avoiding \ion{He}{I} and \ion{Mg}{II} lines 
(Hensberge et al.\ 2000). We compared the observed line profiles to a set of theoretical 
spectra calculated for different $v \sin i$ values. We find $v_{\rm A} \sin i = 98 \pm 2$\kms.
 The small errorbar is due to the very good agreement between the observed and calculated
 line profiles, and also the very high S/N (about 1000) in the disentangled spectrum of star A.

\subsection{Helium abundance} \label{sec:abuhe}

\begin{figure} \centering
\begin{tabular}{cc}
\includegraphics[width=35mm]{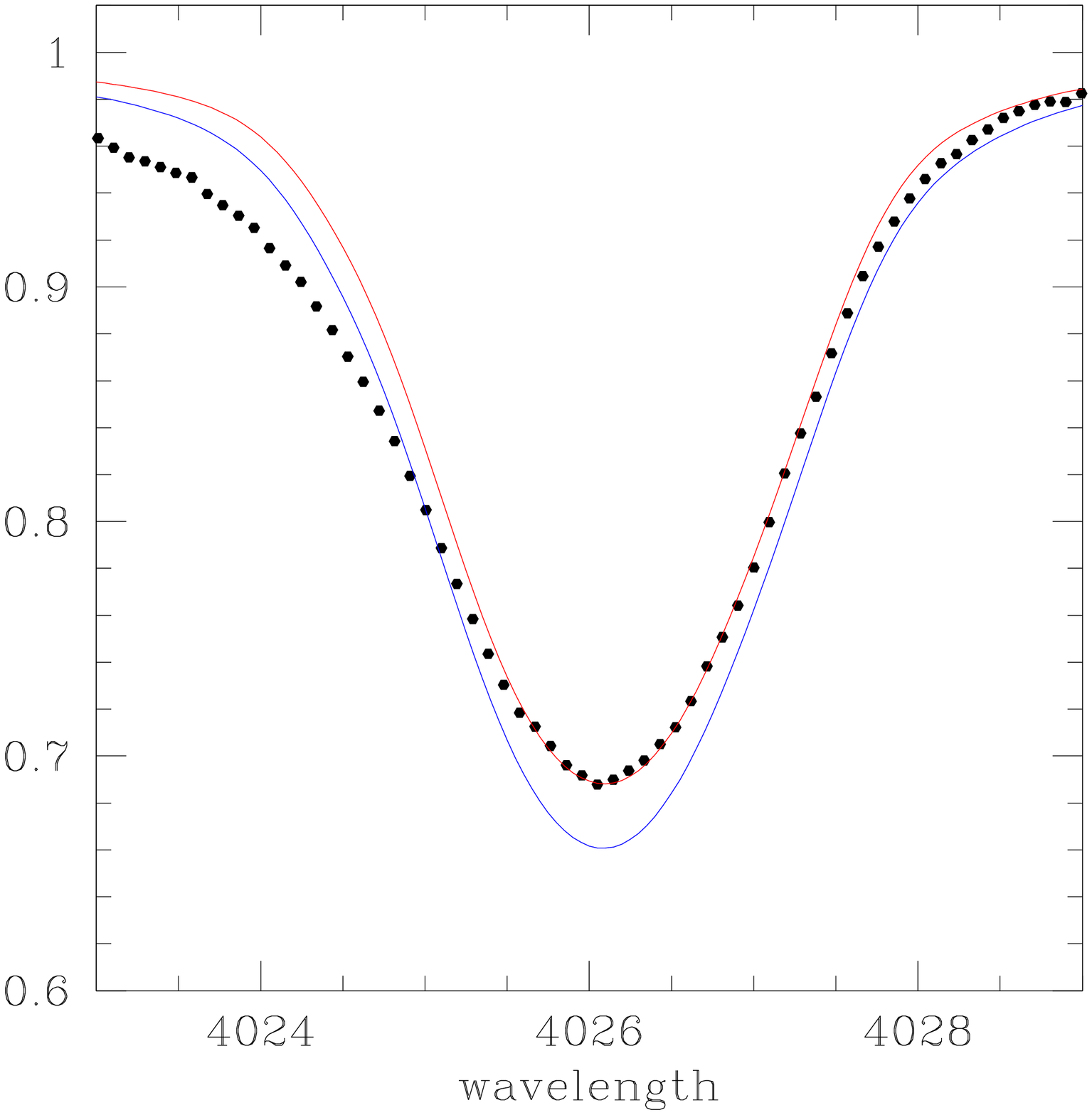} & \includegraphics[width=35mm]{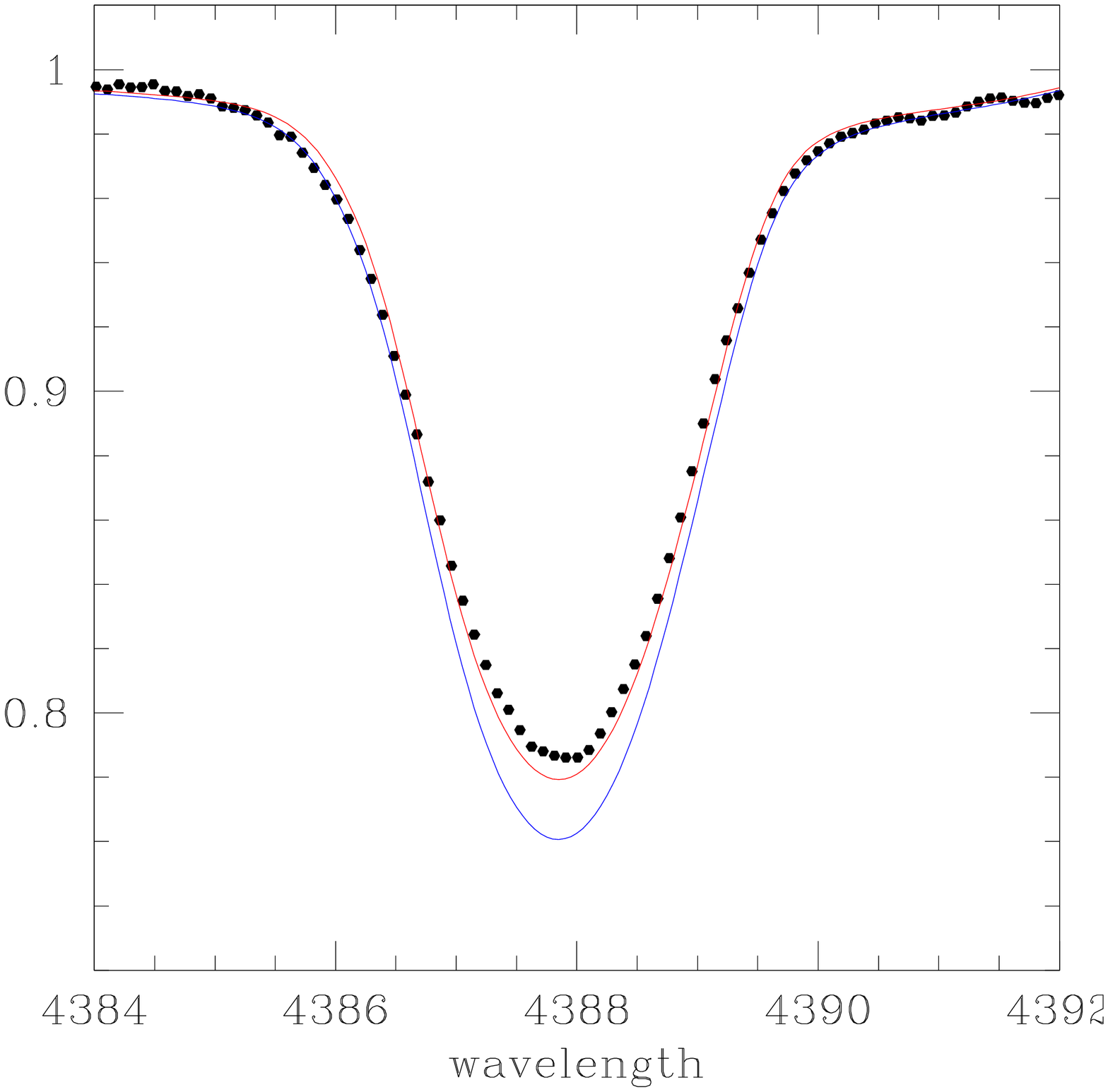} \\
\includegraphics[width=35mm]{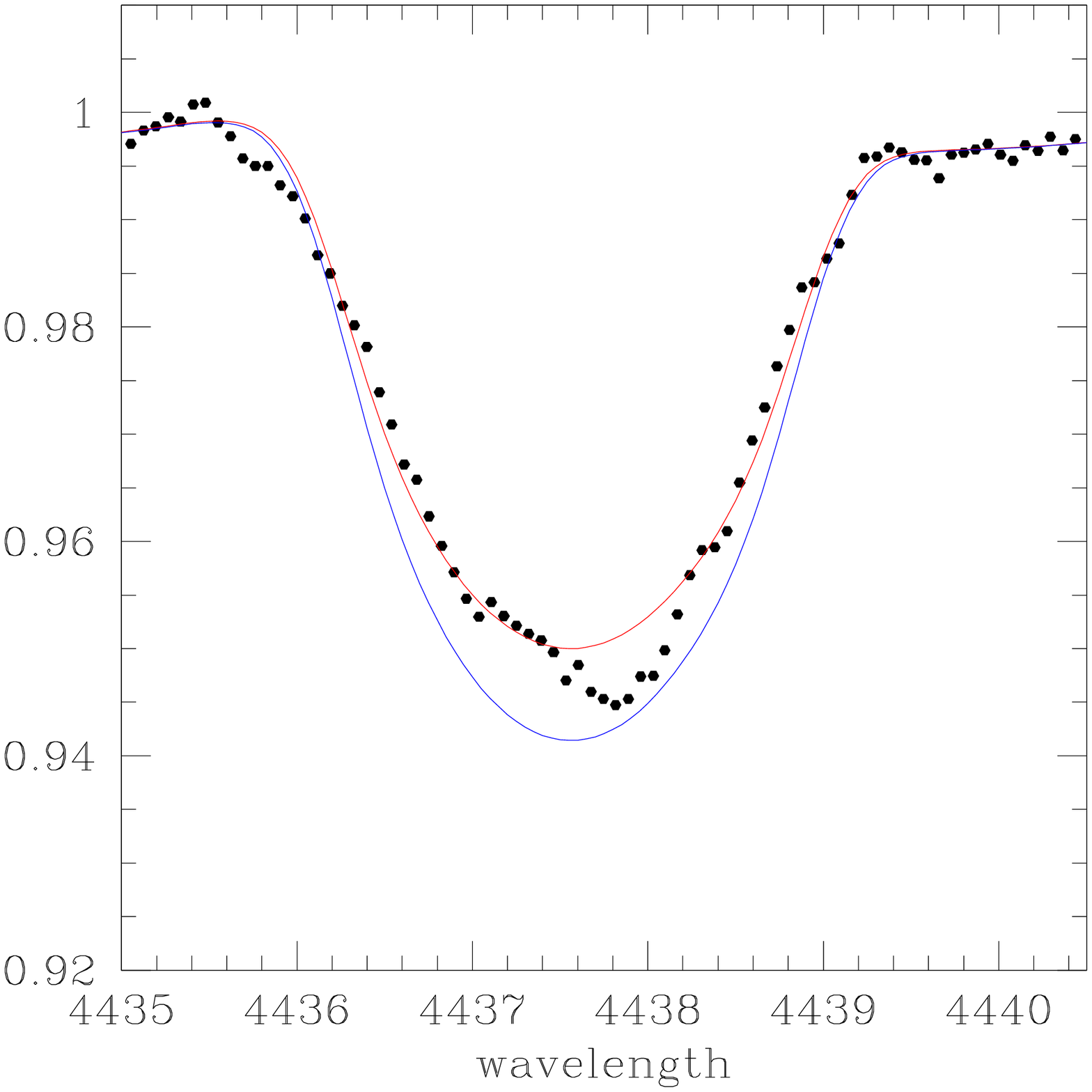} & \includegraphics[width=35mm]{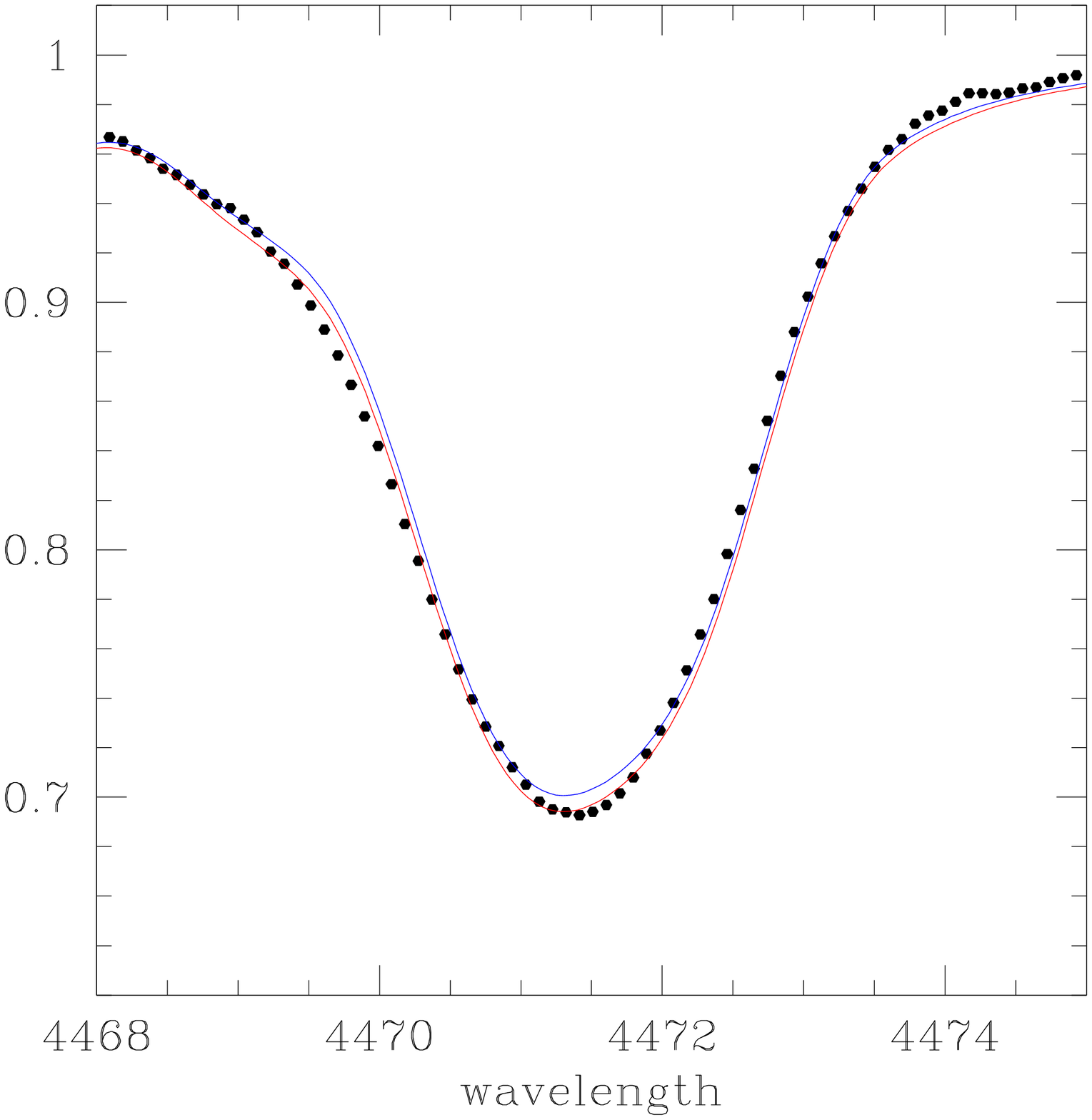} \\
\includegraphics[width=35mm]{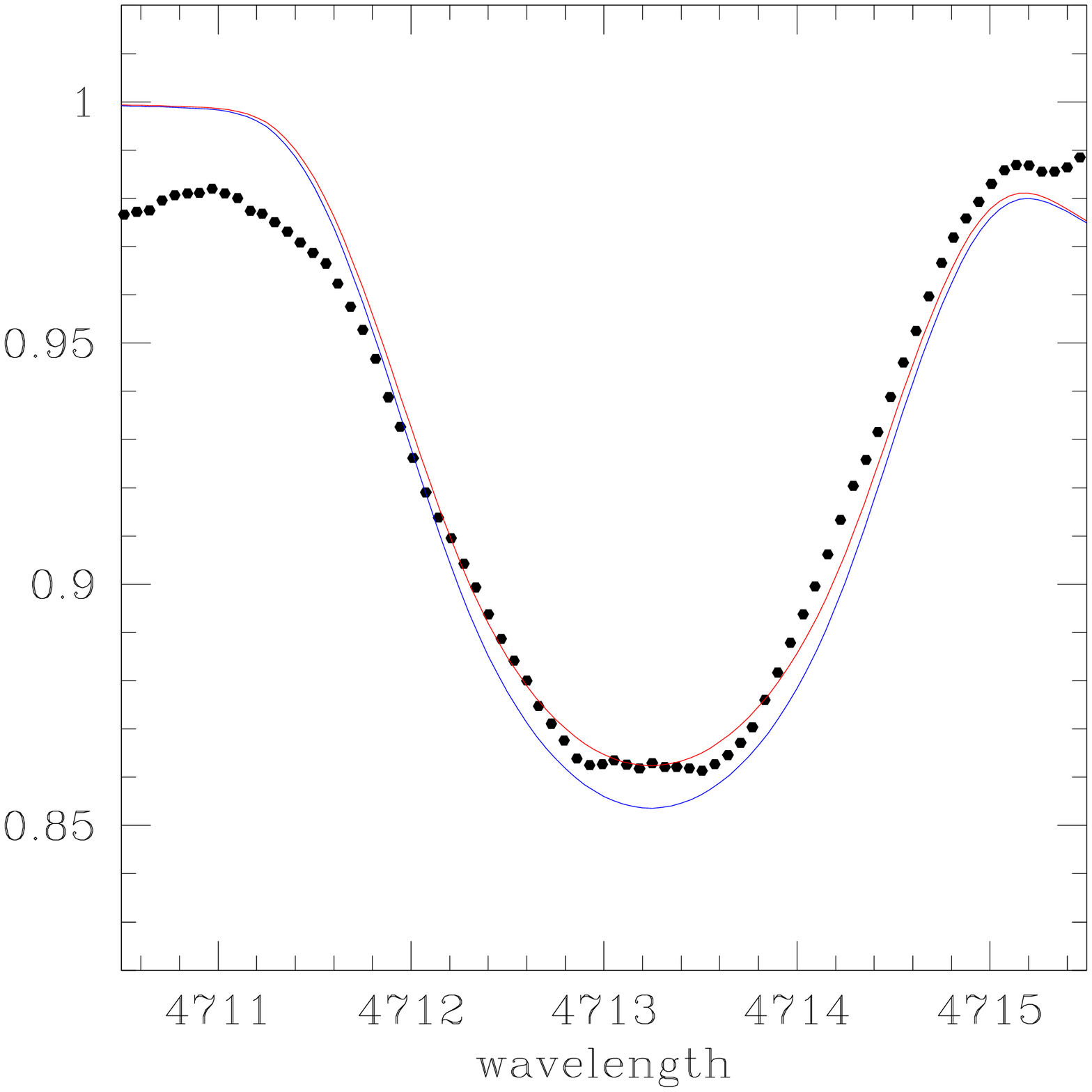} & \includegraphics[width=35mm]{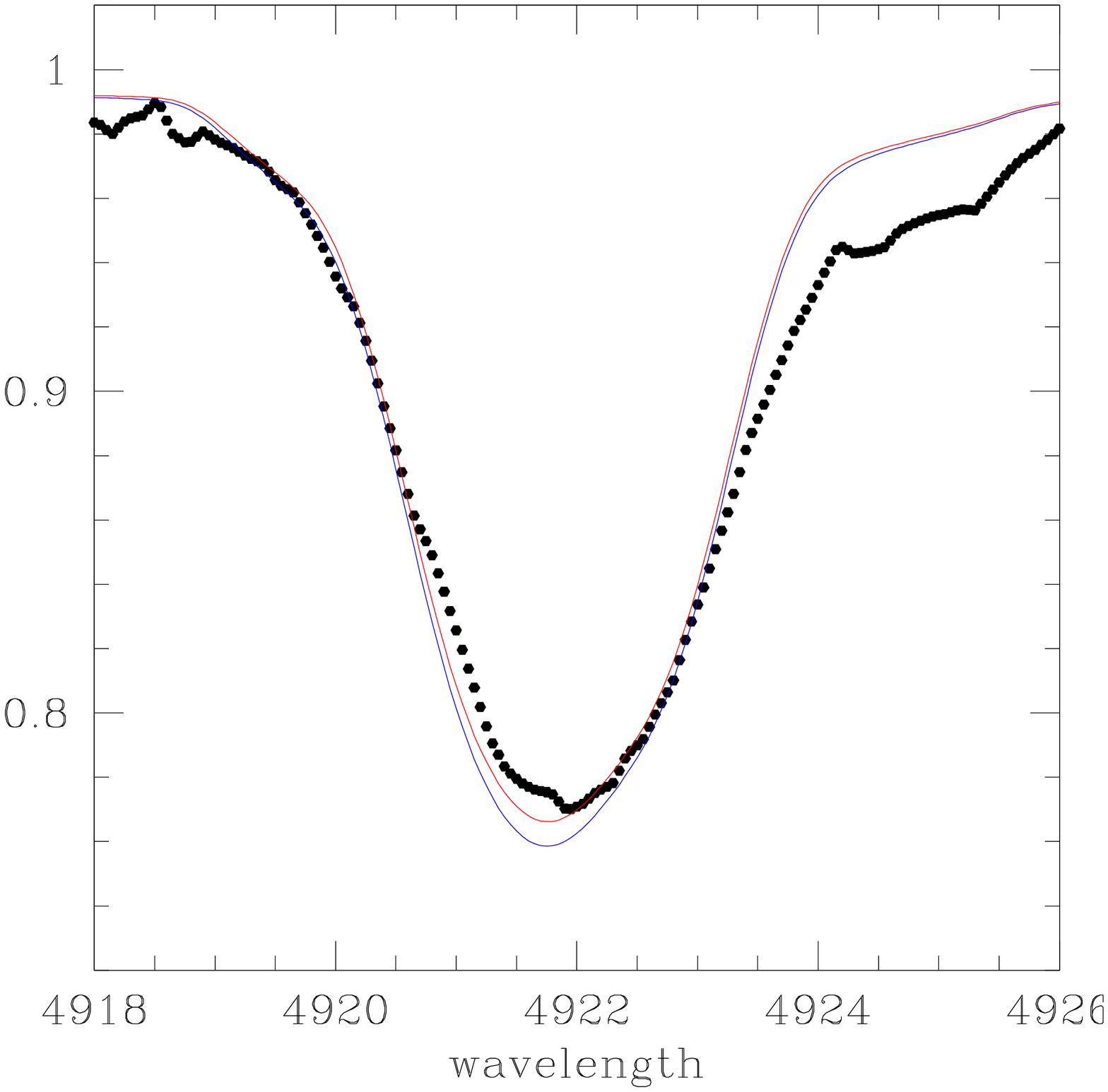} \\
\includegraphics[width=35mm]{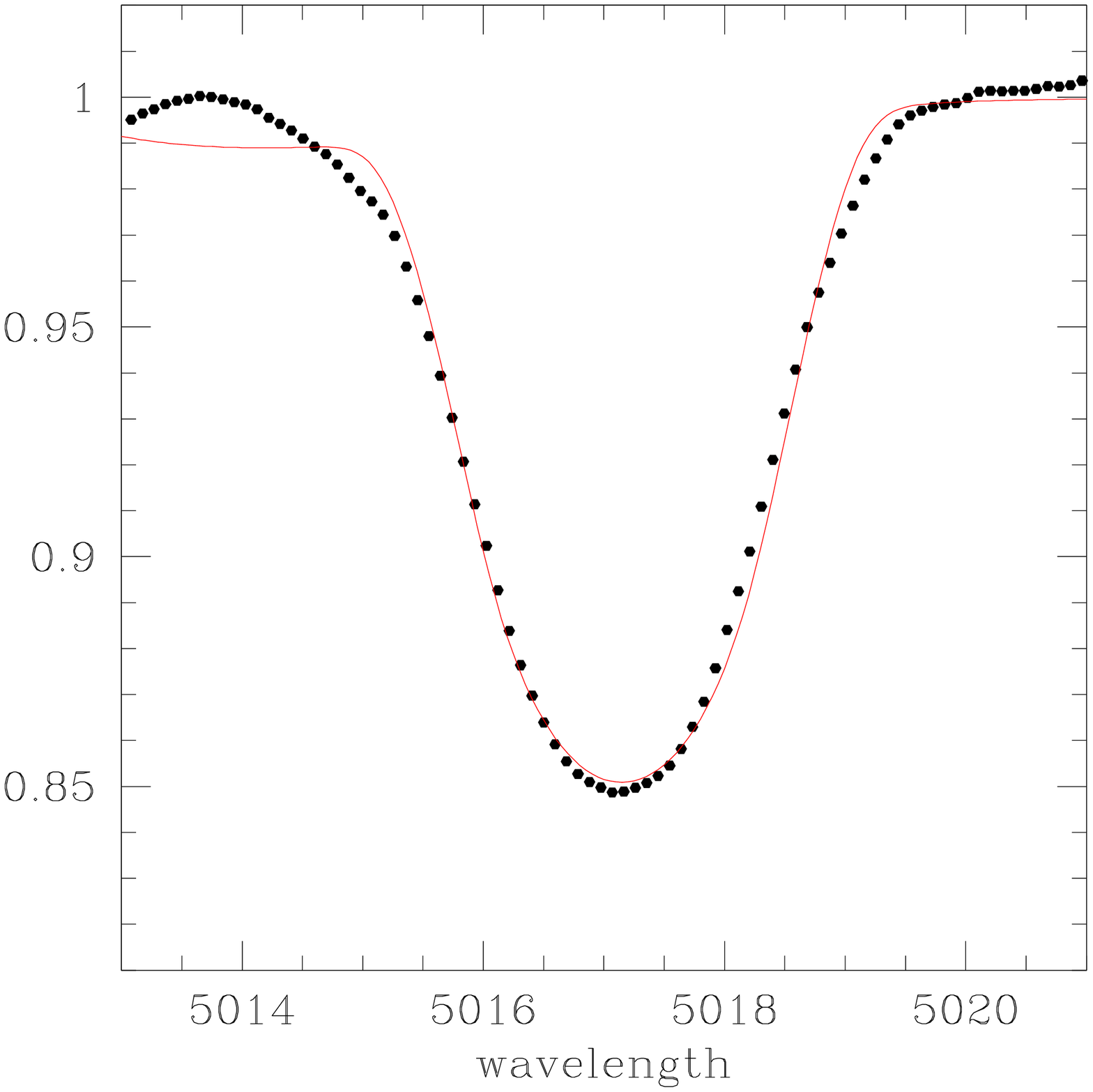} & \includegraphics[width=35mm]{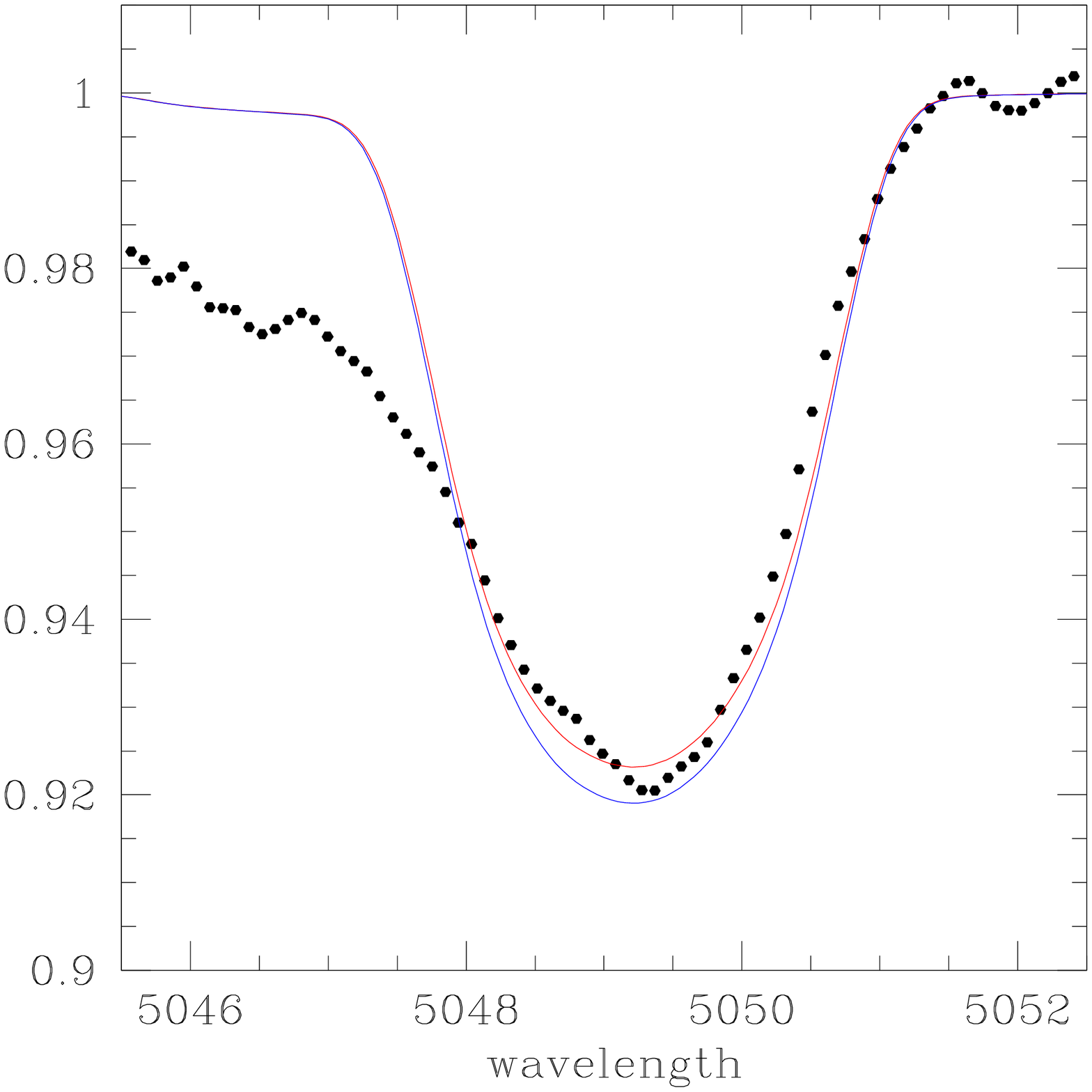} \\
\includegraphics[width=35mm]{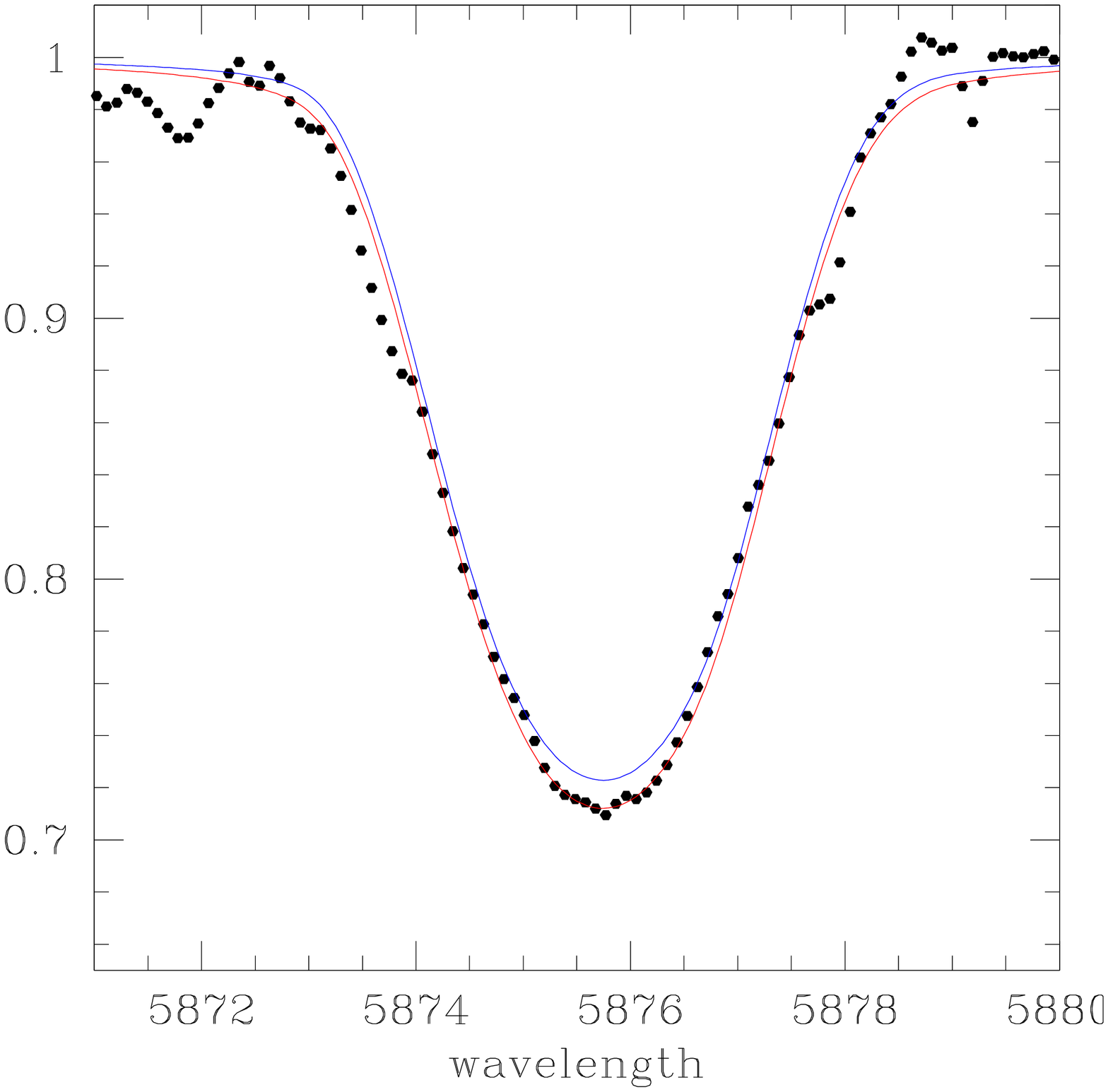} & \includegraphics[width=35mm]{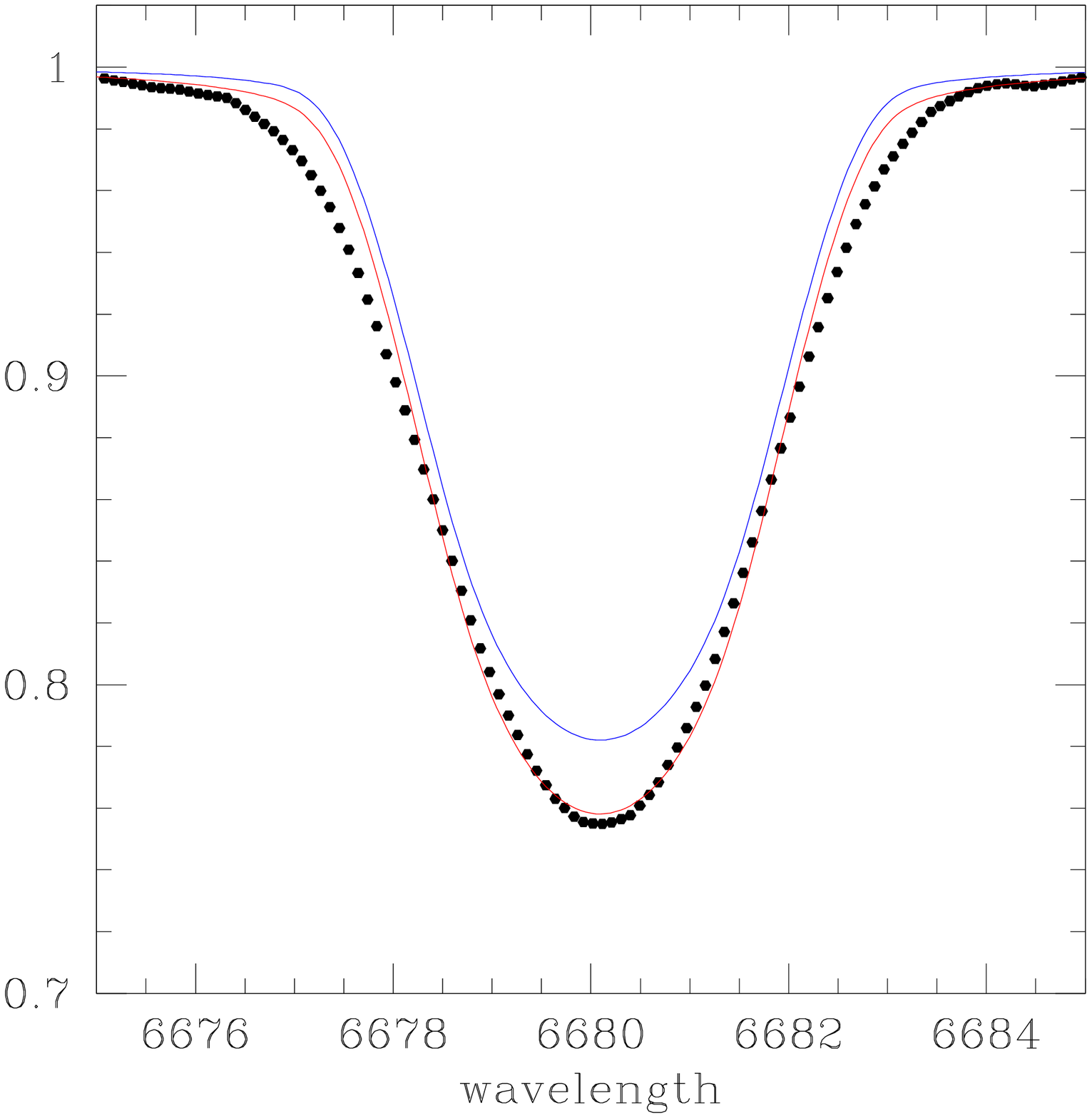} \\
\end{tabular}
\caption{\label{fig:fithel} Profiles of a range of \ion{He}{I} lines. Disentangled spectra
are represented by points, and theoretical line profiles which are the best fit to the
disentangled profiles are shown with red lines. For comparison, profiles calculated for
the solar helium abundance are also plotted with blue lines. The derived abundances are
given in Table\,\ref{tab:abuhe}.}
\end{figure}

\begin{table} \centering \caption{\label{tab:abuhe} Helium abundances
determined for V380\,Cyg\,A from the set of \ion{He}{I} lines, adopting
$\micro = 14$\kms\ as found from the \ion{O}{II} lines. The third column
gives the helium abundances derived by Lyubimkov et al.\ (1996).}
\begin{tabular}{lcc} \hline
Line (\AA)  & $\epsilon$(He)        & $\epsilon$(He)         \\
            & This work             & LRR96             \\
\hline
4026.2      & $0.075 \pm 0.003$     &  -                 \\
4387.9      & $0.069 \pm 0.005$     & 0.123             \\
4437.6      & $0.075 \pm 0.005$     &   -                \\
4471.5      & $0.099 \pm 0.007$     & 0.153             \\
4713.2      & $0.071 \pm 0.008$     & 0.212             \\
4921.9      & $0.078 \pm 0.007$     & 0.145             \\
5015.7      & $0.089 \pm 0.012$     & (0.286)           \\
5047.7      & $0.081 \pm 0.008$     &    -               \\
5875.7      & $0.141 \pm 0.013$     & ($\ge 0.5$)       \\
6678.1      & $0.157 \pm 0.018$     & ($\ge 0.5$)       \\
\hline
mean        & $0.094 \pm 0.031$     & $0.158 \pm 0.038$ \\
\hline \end{tabular} \\
\end{table}

V380\,Cyg\,A has been classified as spectral type B2\,III, in agreement with the \Teff\ and 
\logg\ we find here. In B2 stars lines of neutral helium are very strong, while lines of ionised 
helium are completely absent. We therefore calculated profiles of \ion{He}{I} lines using the 
non-LTE codes {\sc detail} and {\sc surface} (see Paper\,I for details), for helium abundances 
$\epshe = 0.05$--$0.25$  in steps of 0.05. The synthetic spectra were broadened by $v \sin i = 
98$\kms\ and $\micro = 14$\kms, and $\epshe$ was determined by $\chi^2$ minimisation to the 
disentangled and renormalised spectrum. The results for ten \ion{He}{I} lines are given in 
Table\,\ref{tab:abuhe}. The lines of \ion{He}{I} at $\lambda$4009.3 and $\lambda$4120.8 
were not used: our model atom does not include the first line and the second is severely blended 
with \ion{O}{II} $\lambda$4121.1. The resulting mean value for the helium abundance in 
the photosphere of star A is $\epshe = 0.094 \pm 0.031$, which is slightly above the solar value 
[$\epsilon({\rm He})_{\odot} = 0.089$; Grevesse et al.\ (2007)]. The helium abundances found 
from different \ion{He}{I} lines span a range 0.07--0.16 (from subsolar to 80\% above solar 
abundance) and such a large spread needs explaining.

Firstly, the $\lambda$4471 and $\lambda$4922 line calculations are more precise as they include 
transitions up to the $n=4$ level (Przybilla \& Butler 2001). Lyubimkov, Rostopchin \& Lambert 
(2004, hereafter LRL04) discuss this in detail, and also show that the other two diffuse
 \ion{He}{I} lines ($\lambda$4026 and $\lambda$4388) give a systematically lower helium 
abundance. LRL04 selected the $\lambda$4471 and $\lambda$4922 lines as their principal
 helium abundance indicators for B stars. For V380\,Cyg\,A the mean helium abundance 
(by number of atoms) is $N({\rm He}) = 0.072 \pm 0.006$ from $\lambda$4026 and $\lambda$4388
 whereas we find $N({\rm He}) = 0.089 \pm 0.010$ (equalling the solar value) from 
$\lambda$4471 and $\lambda$4922. This corroborates the findings by LRL04.

Secondly, the varying sensitivity of \ion{He}{I} lines to \micro\ might cause abundance 
discrepancies. In this work we found \micro\ from \ion{O}{II} lines, which are the most 
numerous lines in the spectrum of star A. The recent literature contains some discussion 
about whether the \micro\ derived from helium lines should equal that found from metallic
 lines such as \ion{O}{II}, \ion{N}{II} and \ion{Si}{III} (see Hunter et al.\ 2007 and 
references therein). LRL04 found that \micro\ derived from helium lines is systematically 
larger than that derived from metallic lines. Contrary to this, during our work on V453\,Cyg 
(Paper\,I) we simultaneously fitted the helium line profiles for abundance and \micro, 
finding a high \epshe\ and a \micro\ substantially lower than those from \ion{O}{II} lines. 
We rejected this possibility and instead fixed \micro\ to the 14\kms\ measured from the 
\ion{O}{II} lines, which resulted in a normal helium abundance.

The other lines listed in Table\,\ref{tab:abuhe} ($\lambda\lambda$ 4713, 5016, 5048, 5876 
and 6678) are known to have substantial dependences on \micro. Moreover, the last two lines
 are unusually broad when comparing to other helium lines (Fig.\,\ref{fig:fithel}) which
 indicates that other (non-thermal) broadening mechanisms are present. V380\,Cyg\,A has 
evolved to the giant stage, and one could expect differences in the atmospheric velocity 
field compared to dwarf stars. LRL04 have discussed the possibility that asphericity 
and/or mass loss high in the atmospheres of hot B giants could cause excess broadening 
for some helium lines. (Herrero, Puls \& Villamariz 2000). Moreover, they have estimated
 that the $\lambda$5875 and $\lambda$6678 lines would give the largest helium abundances, 
in agreement with our findings. We will discuss this further in Sec.\,\ref{sec:models}.

Thirdly, it is possible that the atmosphere of star A is stratified in helium. Catanzaro 
(2008) has argued that variations in helium abundance derived from different lines can be 
explained by helium stratification. A comparison of his results (for the B2\,V star HD\,32123)
 with ours (Table\,\ref{tab:abuhe}) reveals the same trends from line to line. However, 
a more detailed investigation of this phenomenon will require state-of-the-art non-LTE 
model atmospheres.

\subsection{CNO abundance}                                          \label{sec:abucno}

\begin{table} \centering \caption{\label{tab:abuall} Abundances derived for
V380\,Cyg\,A. The number of lines used are given in the brackets. Abundances
for the solar photosphere and for Galactic OB stars (Hunter et al.\ 2009) are
given in column 3, and column 4 lists solar abundances from Grevesse et al.\
(2007). Abundances for the metallic elements are are given on the scale in
which $N$(He) $=$ 12.}
\begin{tabular}{llll} \hline
Element                   &  Abundance         & OB stars          & Solar         \\
\hline
$N$(He)/$N$(H)            &0.098$\pm$0.010 [10]&0.098$\pm$0.014$^a$&0.093$\pm$0.002\\
$\log \epsilon({\rm C})$  & 8.21$\pm$0.03 [8]  & 8.00$\pm$0.19     & 8.39$\pm$0.05 \\
$\log \epsilon({\rm N})$  & 7.52$\pm$0.10 [24] & 7.62$\pm$0.12     & 7.78$\pm$0.06 \\
$\log \epsilon({\rm O})$  & 8.54$\pm$0.14 [35] & 8.63$\pm$0.16     & 8.66$\pm$0.05 \\
$\log \epsilon({\rm Mg})$ & 7.52$\pm$0.05 [1]  & 7.25$\pm$0.17     & 7.53$\pm$0.09 \\
$\log \epsilon({\rm Si})$ & 7.26$\pm$0.18 [13] & 7.42$\pm$0.07     & 7.51$\pm$0.04 \\
$\log \epsilon({\rm Al})$ & 6.04$\pm$0.03 [4]  & 5.94$\pm$0.14$^b$ & 6.37$\pm$0.04 \\
\hline \end{tabular} \\
\begin{flushleft}
$a$ Derived for unevolved B stars by Lyubimkov et al.\ (2004) \\
$b$ Derived for B stars with $\Teff \le 30\,000$\,K by Daflon et al.\ (2003) \\
\end{flushleft}
\end{table}

\begin{figure} \includegraphics[width=85mm]{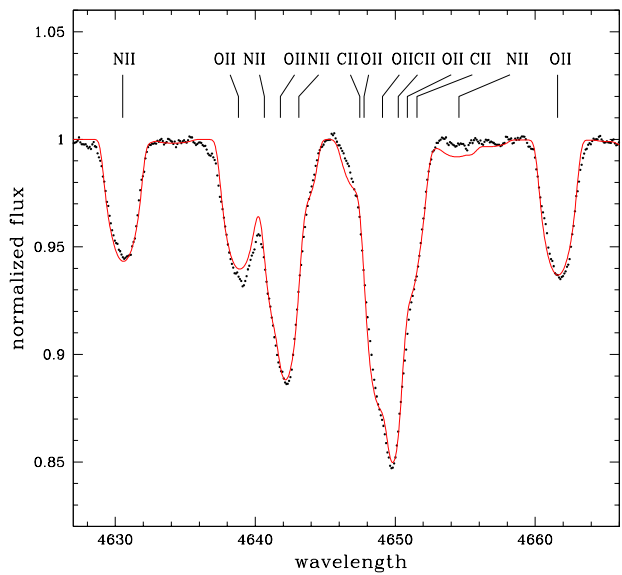}
\caption{\label{fig:plotcno} Comparison between the observed spectrum
of star A (points) and the best-fitting synthetic spectrum (red line)
calculated for the abundances listed in Table\,\ref{tab:abuall}. The
spectral range is 4625--4675\,\AA, which contains heavily blended
lines from \ion{C}{II}, \ion{N}{II} and \ion{O}{II}.} \end{figure}

The surface abundances of helium and the CNO elements are important observational results for 
comparison with theoretical predictions, because models including rotational effects predict 
an enrichment of He and N and a depletion of C and O during the evolution of B stars. 
The optical spectra of these objects are rich in O and N lines, but have fewer C lines. 
We determined \micro\ and the oxygen abundance in Sec.\,\ref{sec:micro}, by fitting synthetic
 spectra to the observed disentangled spectra of star A. Using the same approach for C and N 
gives results which show relatively low scatter between lines. The model carbon atom we used 
cannot reprode the strongest line (\ion{C}{II} $\lambda$4267), but the abundance estimates 
for the fairly strong doublet on the red wing of H$\alpha$ ($\lambda$6556 and $\lambda$6559) 
are consistent with the abundances from other C lines.

In the spectral range 4635--4655\,\AA\ there is a bunch of relatively strong \ion{O}{II}, 
\ion{C}{II} and \ion{N}{II} lines, which can provide reliable estimates and checks on the 
derived abundances for these elements (Pavlovski \& Hensberge 2005). In Fig.\,\ref{fig:plotcno} 
we show that there is good agreement between the disentangled spectrum of star A and a theoretical
 spectrum calculated for the abundances derived from unblended lines (Table\,\ref{tab:abuall}).

N is more sensitive to changes in abundances than C, for which only minute changes are predicted.
 However, a large spread in N abundances have been found for MS B stars, and mechanisms other 
than rotational mixing may be responsible (Morel et al.\ 2006, Hunter et al.\ 2009). This is
 highlighted by the [N/C] ratio, which is a robust indicator of CNO-processed material dredged 
up to the stellar surface. Its distribution in B stars is clearly bimodal (Morel 2008), with 
one subgroup around the solar value ([N/C]$_\odot  \sim -0.6$\,dex), and one with significantly 
higher values, ([N/C] $\sim -0.1$\,dex). With [N/C]$_{\rm A} = -0.68 \pm 0.10$\,dex, 
V380\,Cyg\,A is close to the solar value. This is also true for the [N/O] ratio 
([N/O]$_{\rm A} = -1.01 \pm 0.17$\,dex and [N/O]$_\odot = -0.90$\,dex).

\subsection{Magnesium, silicon and aluminium}                                                        \label{sec: abumag}

\begin{figure} \centering \includegraphics[width=70mm]{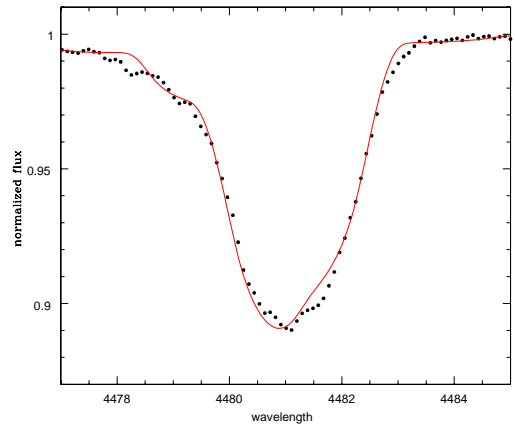}
\caption{\label{fig:alumag} The best-fitting calculated profiles
(solid line) of \ion{Mg}{II} $\lambda$4481.2, which is blended
with the much weaker \ion{Al}{III} $\lambda$4479.9 line, compared
to the observed profiles (filled symbols) for star A.} \end{figure}

Determination of the abundances of metals like Mg, Si and Al makes possible an estimate of 
the bulk metallicity of B stars. These elements do not participate in the CNO cycle so their 
abundances are not affected by MS evolution. 
%%Also, it is not expected that the surface Mg 
%%and Al have been altered by the mixing of material exposed in the Mg-Al cycle (Daflon et al.\ 
%%2003). 
In particular, Mg is an excellent metallicity indicator because its solar and meteoritic
 abundances are know with high accuracy. 
%%For the solar magnesium abundance Holweger (2001) 
%%derived $\log \epsilon{\rm (Mg)} = 7.54 \pm 0.06$ from photospheric lines.

A disadvantage in the use of Mg for metallicity estimates in B stars is that there is only 
one suitable spectral line in the whole optical region: \ion{Mg}{II} $\lambda$4481.2; although 
this is one of the strongest lines in the spectra of early-type stars. For stars with moderate 
and high $v\sin i$ an analysis of this line is complicated by blending with the neighbouring 
\ion{Al}{III} $\lambda$\,4479.9 line. In the \Teff\ range of interest here, the \ion{Al}{III} 
$\lambda$4479.9 line reaches its maximum strength, and the ratio of the \ion{Al}{III} and 
\ion{Mg}{II} line EWs reaches a maximum (see Lyubimkov et al.\ 2005).

We derived the Al abundance for star A from three lines ($\lambda\lambda$ 4512, 4671, 5766), 
finding a very good agreement. We used this measurement in calculating the Mg abundance from 
the \ion{Al}{III} and \ion{Mg}{II} blend at 4481\,\AA. We find $\log \epsilon{\rm (Mg)} = 
7.52 \pm 0.05$, which translates to a metallicity index of [Mg/H] $= -0.02 \pm 0.05$. 
Fig.\,\ref{fig:alumag} shows the fit of the synthetic spectrum to the data for this 
spectral line blend.

%%In their extensive study of elemental abundances in B stars, Lyubimkov et al.\ (2005) adopted 
%%the value $\log \epsilon{\rm (Mg)} = 7.59 \pm 0.15$ as the final mean Mg abundance in B stars, 
%%which gives [Mg/H] $= 0.04 \pm 0.15$. Our value for star A complements their result, and 
%%indicates that its bulk metallicity is approximately solar.

G2000 found a subsolar metallicity for star A ([M/H] $= -0.44 \pm 0.07$) by fitting the 
system's spectral energy distribution. Elemental abundances in the present work are derived 
from non-LTE line profile calculations, while G2000 used theoretical spectra generated 
in the LTE approximation. This might explain the disagreement between the two results.

\subsection{Secondary star}

\begin{figure} \centering
\includegraphics[width=70mm]{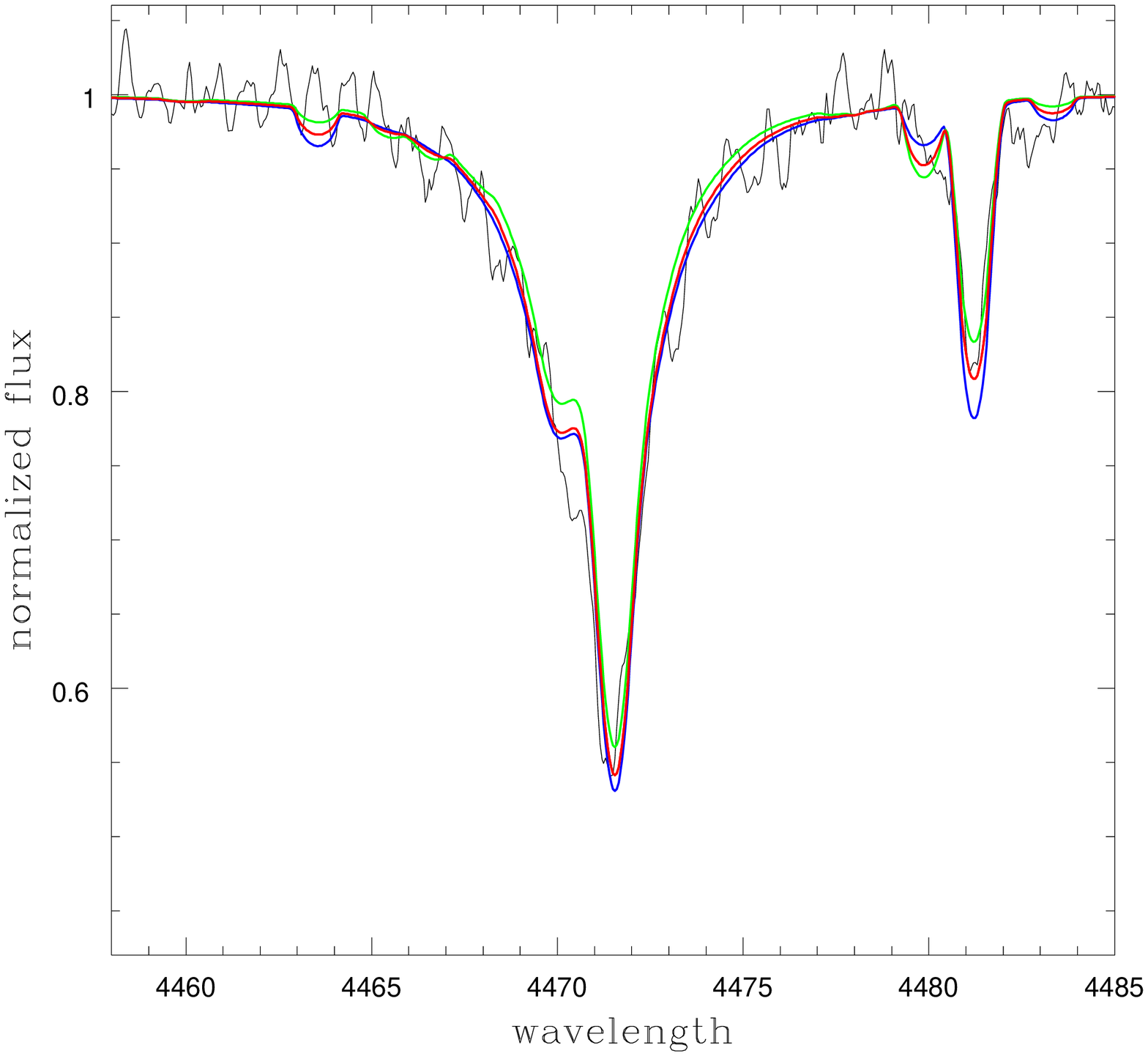} \
\includegraphics[width=70mm]{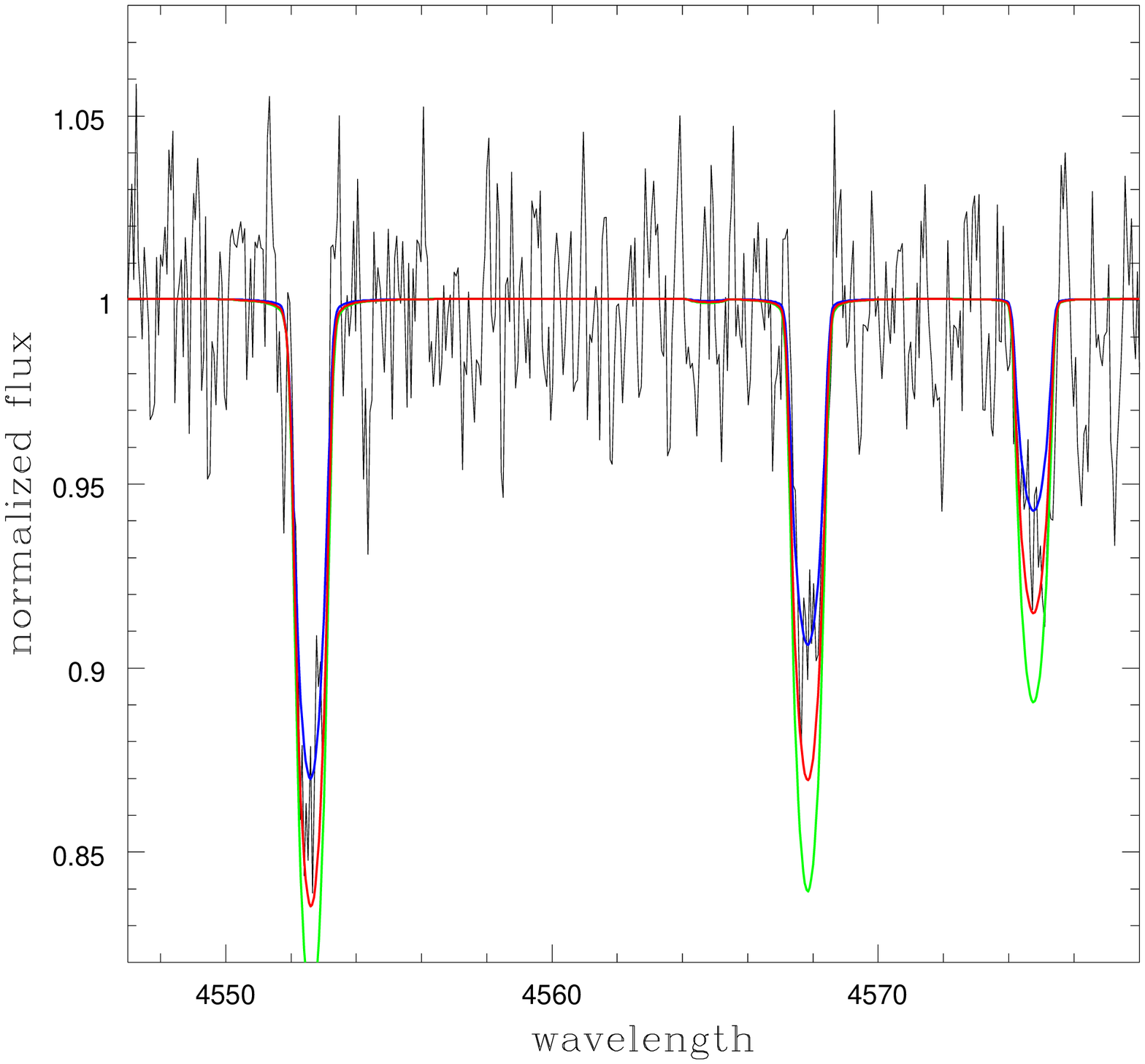} \\
\caption{\label{fig:secspe} Comparison of the disentangled renormalised
spectrum of star B to theoretical spectra for two wavelength intervals,
covering \ion{He}{I} $\lambda$4471 and \ion{Mg}{II} $\lambda$4481 and
the \ion{Si}{III} triplet at $\lambda\lambda$4550-4580. The synthetic
spectra were calculated for \Teff = 20\,000 K (blue lines), 22\,000 K
(red) and 24\,000 K (green). Elemental abundances were the same as for
star A. The low S/N of the disentangled spectrum is due to relative
faintness of star B.} \end{figure}

Despite its faintness and small contribution to the total flux of the system, our 
application of {\spd} readily revealed the spectrum of star B (Fig.\,\ref{fig:plotspe}). 
Its light contribution, determined from the light curve analysis (Sec.\,\ref{sec:lc}), 
is 6.1\% in $B$ and 5.9\% in $V$. To renormalise its disentangled spectrum to a continuum 
level of unity we have multiplied it by 16.4 and 16.9, respectively. With typically 55 
spectra available for disentangling for a given spectral range, with average S/N $\sim$ 150, 
the reconstructed spectrum of star B has S/N $\sim$ 45. This is enough to check its \Teff, 
but too low for a detailed abundance study.

Theoretical spectra were synthesised with $\Teff = 19\,000$--$25\,000$\,K and $\logg = 4.112$.
 Only strong \ion{He}{I} lines, \ion{Mg}{II} $\lambda$4481 and the \ion{Si}{III} 
$\lambda\lambda$4550--4580 triplet were considered in the $\chi^2$ minimisation. 
The mean value from eight spectral lines gives ${\Teff}_{\rm B} = 21\,600 \pm 550$\,K, 
which is consistent within the uncertainties with ${\Teff}_{\rm B}$ and the \Teff\ ratio 
inferred from the light curve analysis.

In Fig.\,\ref{fig:secspe} we compare the disentangled spectrum of star B to theoretical 
spectra for two wavelength intervals, covering \ion{He}{I} $\lambda$4471 and \ion{Mg}{II}
 $\lambda$4481 and the \ion{Si}{III} triplet. The synthetic spectra were generated using 
the He, Mg and Si abundances found for star A. Further progress would require a substantial
 increase in the S/N, and thus extensive new observations.

%%%%%%%%%%%%%%%%%%%%%%%%%%%%%%%%%%%%%%%%%%%%%%%%%%%%%%%%%%%%%%%%%%%%%%%%%%%%%%%%%%%%%%%%%%%%%%%%%%%%%%%%%%%%%%%%%%

\section{The physical properties of V380\,Cyg}                           \label{sec:absdim}

\begin{table} \centering \caption{\label{tab:dimensions} The absolute
dimensions and related quantities determined for V380\,Cyg. \Veq\ and
\Vsync\ are the observed equatorial and calculated synchronous rotational
velocities, respectively. We adopt $\Lsun = 3.826 \times 10^{26}$\,W and
$M_{\rm bol,\odot} = 4.75$.}
\begin{tabular}{l r@{\,$\pm$\,}l r@{\,$\pm$\,}l} \hline
                                    &      \mc{Star A}      &      \mc{Star B}      \\
\hline
Semimajor axis (\Rsun)              & \multicolumn{4}{c}{$62.17 \pm 0.32$}          \\
Mass (\Msun)                        & 13.13     & 0.24      & 7.779     & 0.095     \\
Radius (\Rsun)                      & 16.22     & 0.26      & 4.060     & 0.084     \\
\logg\ (\cmss)                      & 3.136     & 0.014     & 4.112     & 0.017     \\
Effective temperature (K)           & 21\,750   & 280       & 21\,600   & 550       \\
$\log (L / L_\odot)$                & 4.73      & 0.028     & 3.51      & 0.040     \\
$M_{\rm bol}$ (mag)                 & $-$7.06   & 0.06      & $-$4.03   & 0.10      \\
\Veq\ (\kms)                        & 98        & 2         & 43        & 4         \\
\Vsync\ (\kms)                      & 66.1      & 1.1       & 16.54     & 0.34      \\[2pt]
Extinction \EBV\ (mag)              & \multicolumn{4}{c}{$0.20 \pm 0.03$}           \\
\multicolumn{3}{l}{Distance using Flower  $V$-band BCs (pc)}& \mc{$1015 \pm 56$}    \\
\multicolumn{3}{l}{Distance using Bessell $V$-band BCs (pc)}& \mc{$1034 \pm 50$}    \\
\multicolumn{3}{l}{Distance using Bessell $K$-band BCs (pc)}& \mc{$1004 \pm 23$}    \\
\multicolumn{3}{l}{Distance using Girardi $V$-band BCs (pc)}& \mc{$1035 \pm 50$}    \\
% \multicolumn{3}{l}{Distance using Girardi $J$-band BCs (pc)}& \mc{$1058 \pm 25$}    \\
% \multicolumn{3}{l}{Distance using Girardi $H$-band BCs (pc)}& \mc{$1028 \pm 23$}    \\
\multicolumn{3}{l}{Distance using Girardi $K$-band BCs (pc)}& \mc{$1004 \pm 23$}    \\
\hline \end{tabular} \end{table}

Armed with the velocity amplitudes determined in Sec.\,\ref{sec:orbits} and the results 
of the light curve analysis from Sec.\,\ref{sec:lc} we can determine the physical 
properties of the component stars of V380\,Cyg (Table\,\ref{tab:dimensions}). For this 
we use the {\sc jktabsdim} code (Southworth et al.\ 2005), which propagates the 
uncertainties on the input quantities via a perturbation analysis. The resulting 
error budget is dominated by the photometric parameters, highlighting the importance 
of obtaining a new light curve for this difficult object. Even without this, the 
masses and radii of the component stars are now known to accuracies of 2\% or 
better so are good enough for detailed tests of stellar evolution theory. We also 
find that V380\,Cyg has not attained either orbital circularisation or rotational 
synchronisation.

\begin{figure*} \centering \setlength{\tabcolsep}{0pt}
\hspace*{-15pt}\begin{tabular}{ccc}
\includegraphics[width=60mm]{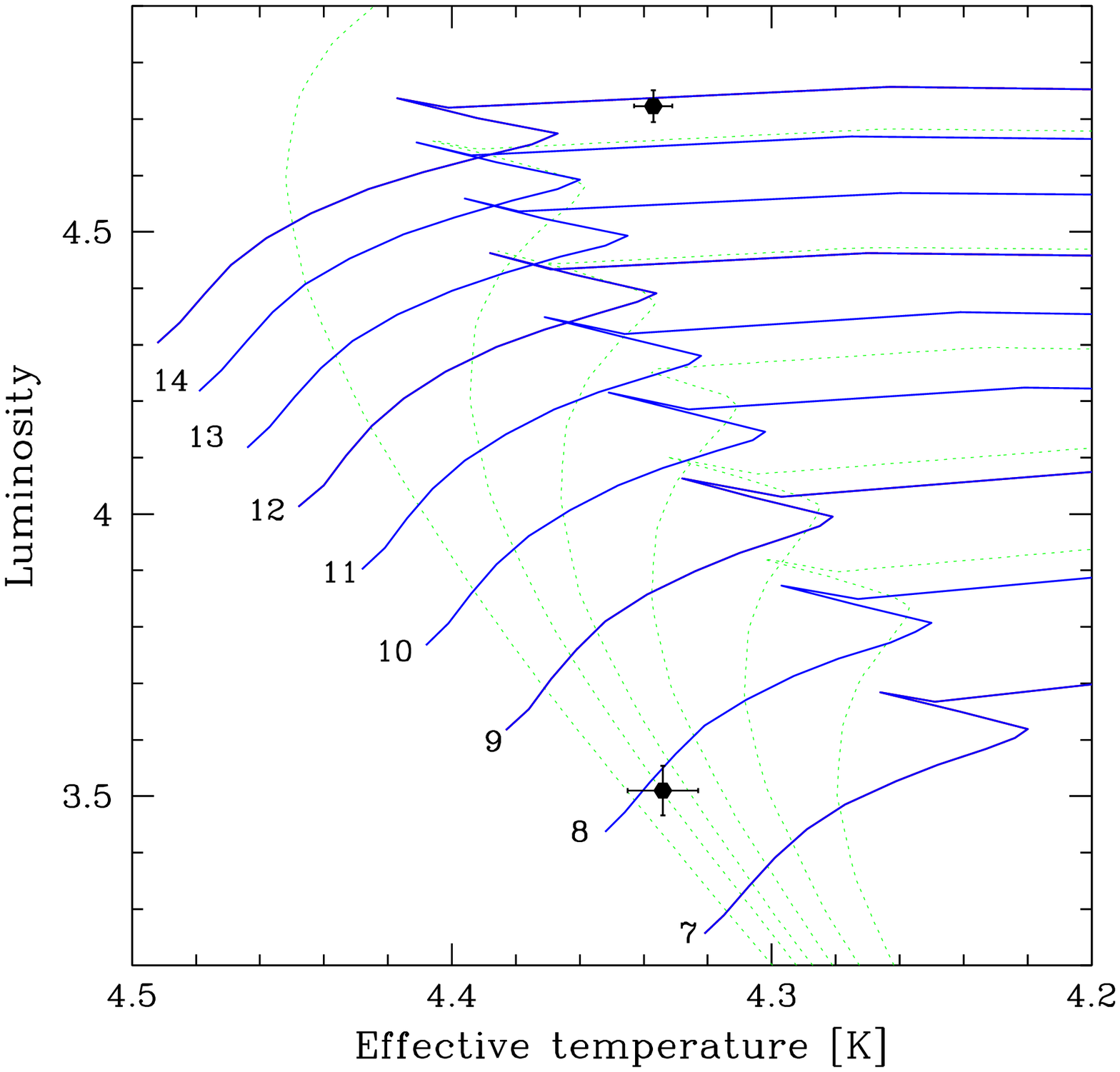} &  \includegraphics[width=60mm]{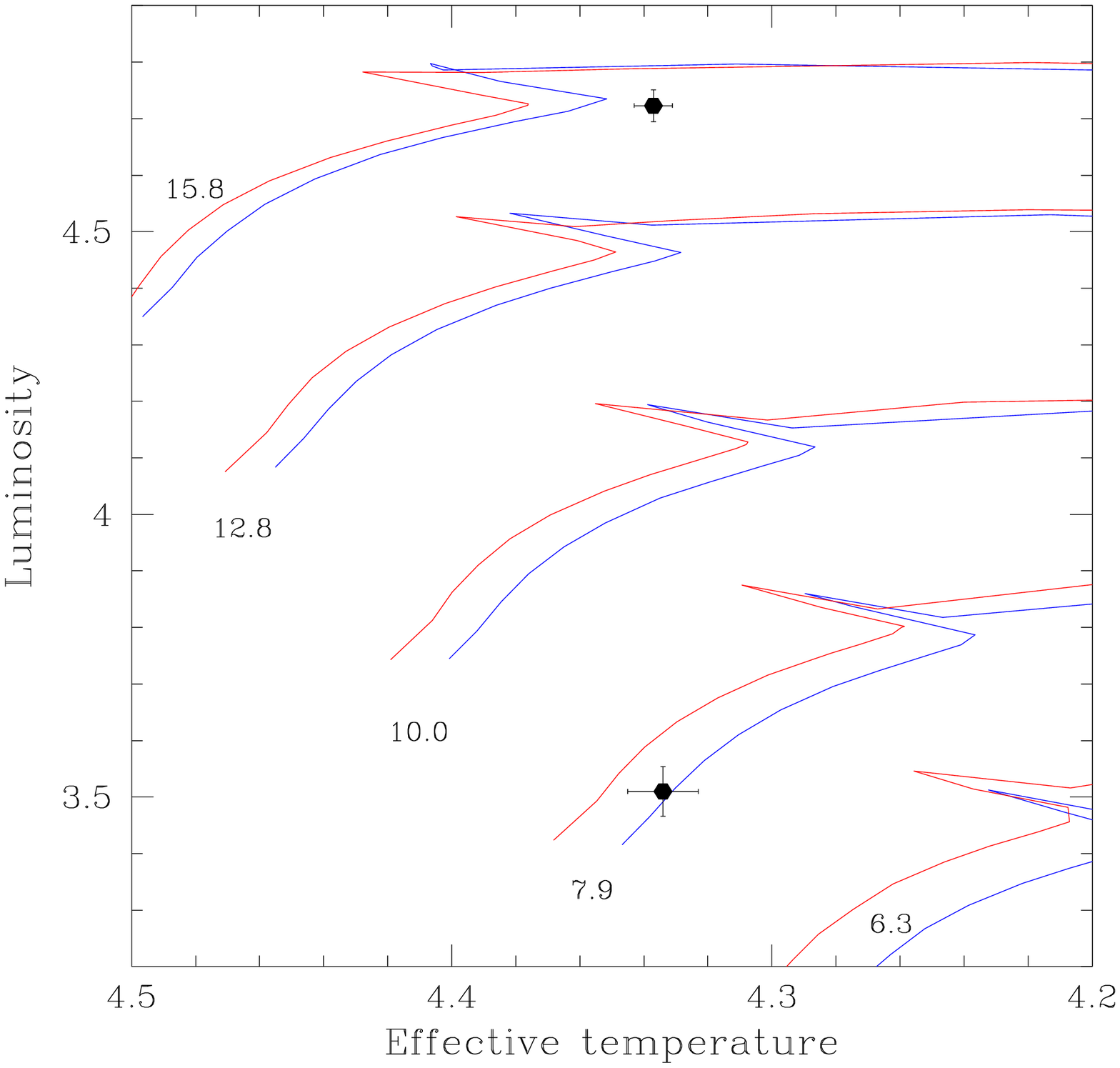} &
\includegraphics[width=60mm]{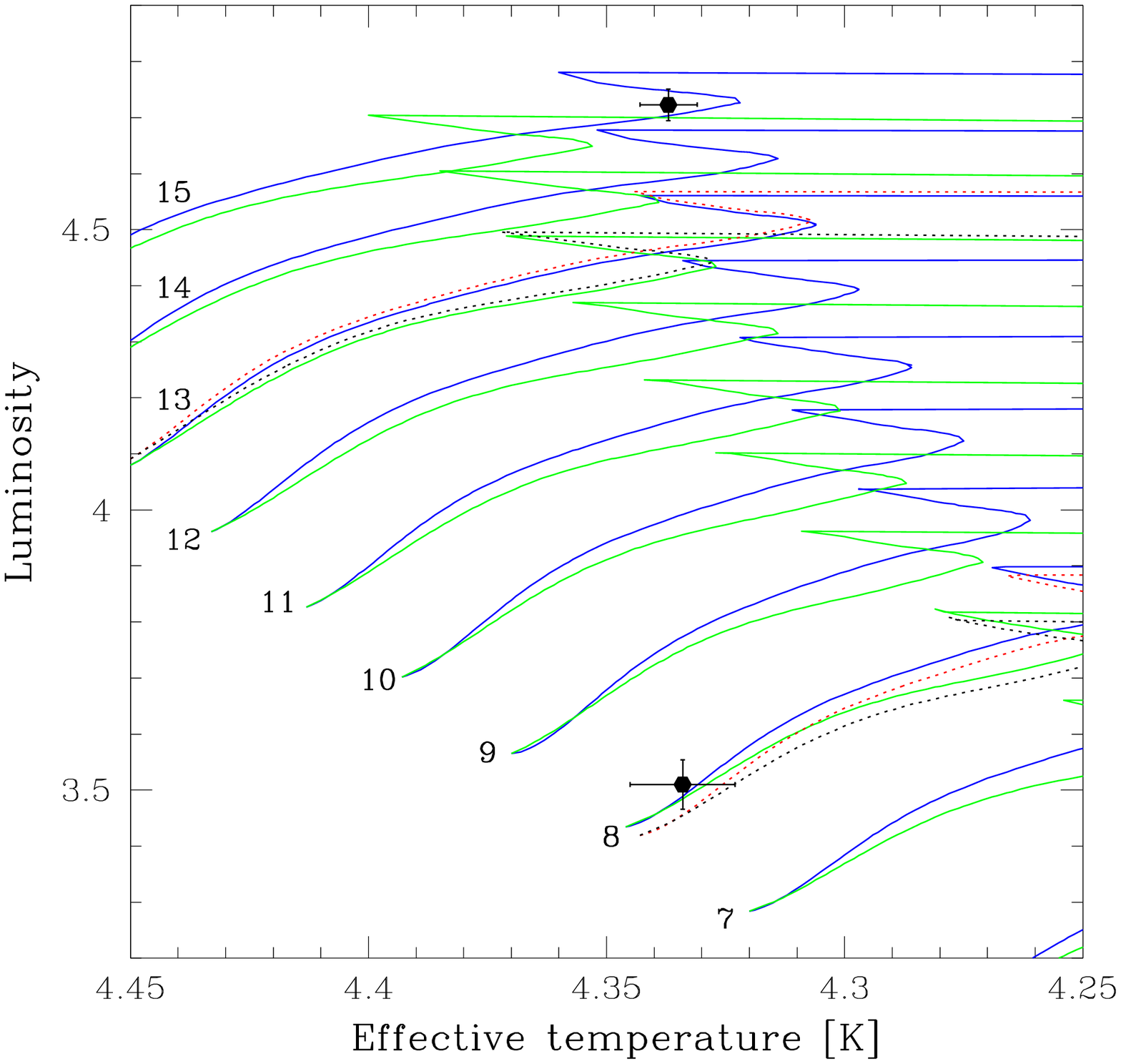}  \\ \includegraphics[width=60mm]{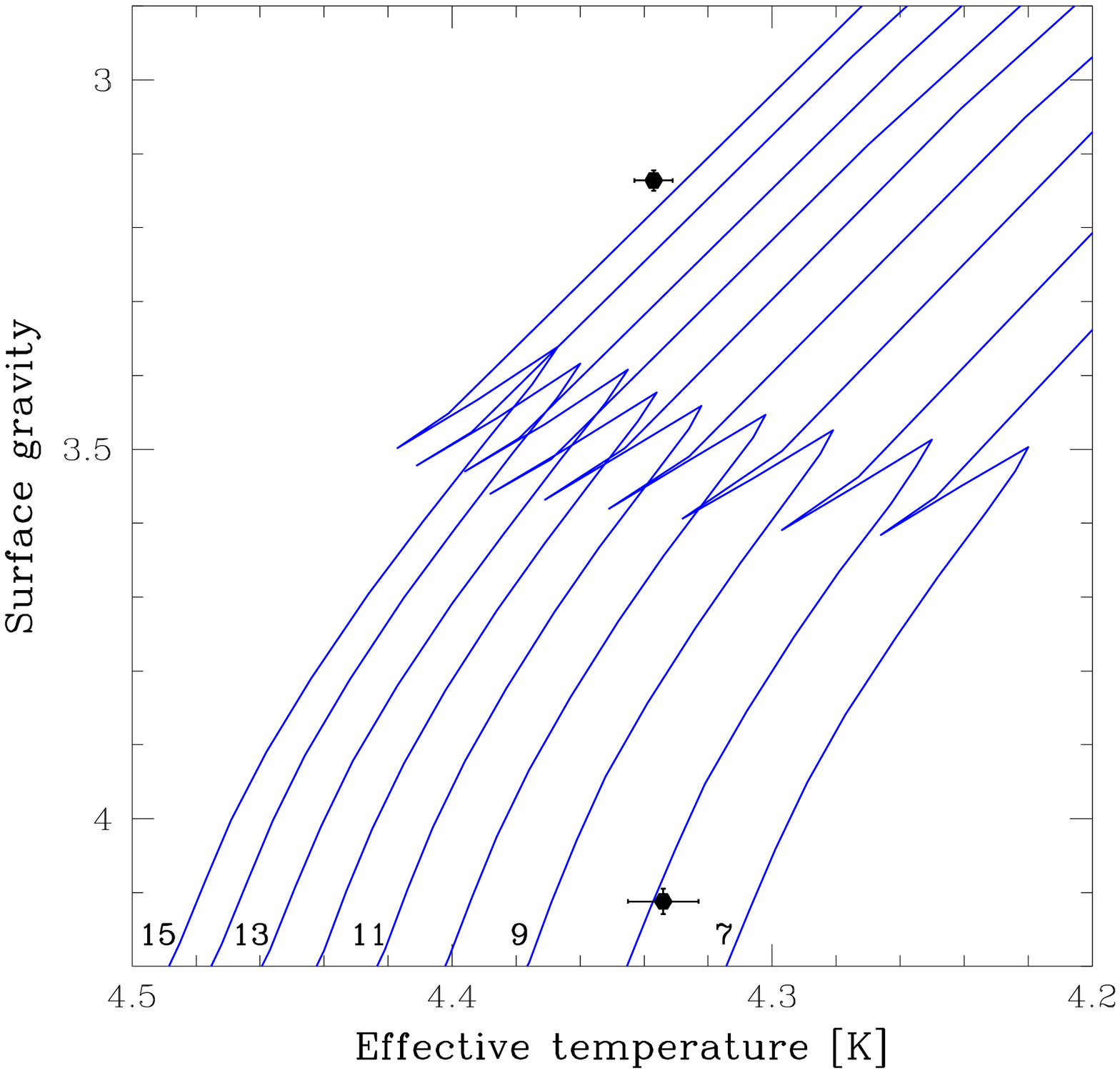}  &
\includegraphics[width=60mm]{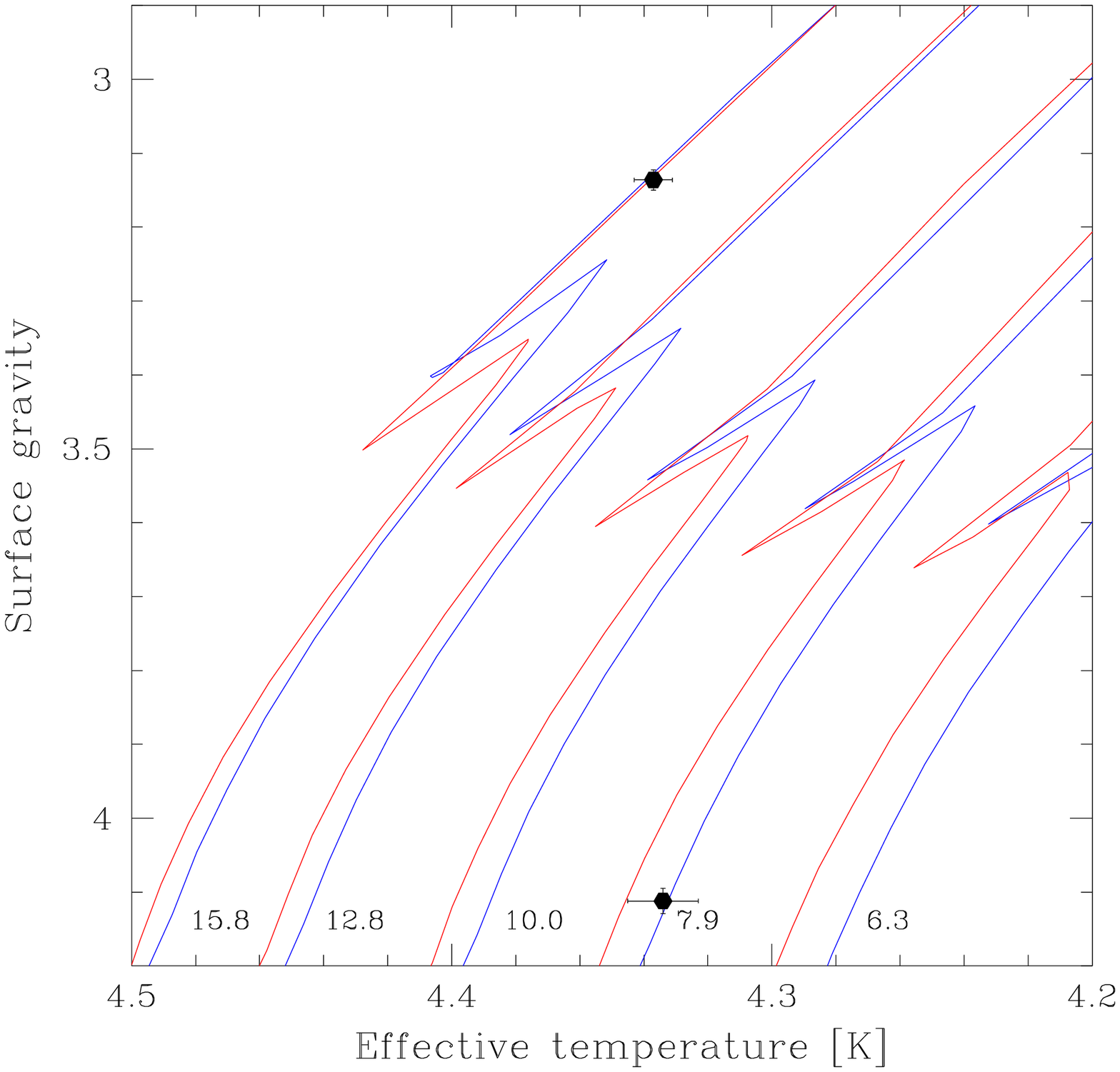} & \includegraphics[width=60mm]{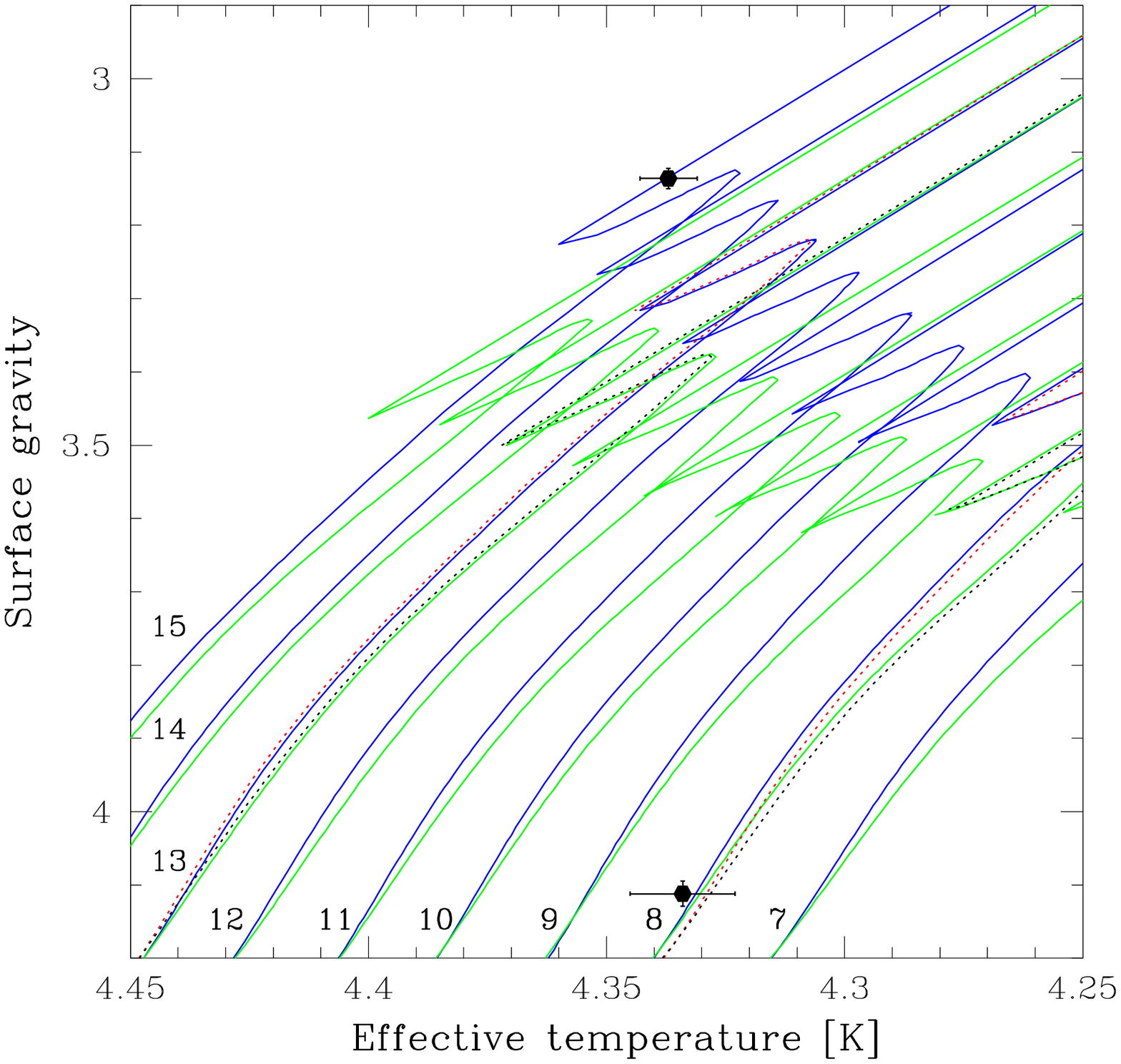} \\
\end{tabular}
\caption{\label{fig:plotshaller}\label{fig:plotrot} Comparison between stellar models and
the absolute parameters of V380\,Cyg in the $\log \Teff - \log L$ and $\log \Teff - \log g$
diagrams. Numbers in the plots give the masses (in \Msun) for the theoretical model tracks,
and the positions of the components of V380\,Cyg are indicated with filled circles.
{\it Left panels:} models calculated with core overshooting but no rotational mixing
(Schaller et al.\ 1992). {\it Centre panels:} models calculated by Claret (1995) for
metallicity $Z=0.02$ (blue lines) and Claret \& Gim\'enez (1995) for $Z=0.01$ (red lines),
with core overshooting but no rotational mixing. {\em Right panels:} models which include
rotational mixing (Ekstr\"{o}m et al.\ 2008), with $\Omega/\Omega_{\rm crit} = 0.10$ (green
lines), $\Omega/\Omega_{\rm crit} = 0.50$ (blue lines). Dotted red and black lines denote
model predictions for the masses of V380\,Cyg (13.1 and 7.8 \Msun) for the same two
rotational velocities.}
\end{figure*}

With the \Teff s determined above, we have found the distance to V380\,Cyg via the 
bolometric correction (BC) method. Empirical BCs were obtained from Flower (1996), and 
theoretical ones from Bessell et al.\ (1998) and Girardi et al.\ (2002). We adopted 
$BV$ magnitudes from the Tycho catalogue (H{\o}g et al.\ 1997) and $JHK$ magnitudes 
from 2MASS (Skrutskie et al.\ 2006). There is good agreement between different sources 
of BCs (Table\,\ref{tab:dimensions}). Using the Girardi et al.\ (2002) BCs, which are 
available for a wide range of passbands, we find that a reddening of $\EBV = 0.20 
\pm 0.03$ is needed to obtain consistent distances at optical and infrared wavelengths. 
For our final value we adopt a distance of $1004 \pm 23$\,pc, from $K$-band BCs. This is 
in good agreement with but more precise than the value of $1000 \pm 40$\,pc found by G2000.

%%%%%%%%%%%%%%%%%%%%%%%%%%%%%%%%%%%%%%%%%%%%%%%%%%%%%%%%%%%%%%%%%%%%%%%%%%%%%%%%%%%%%%%%%%%%%%%%%%%%%%%%%%%%%%%%%%

\section{Probing the models}   \label{sec:models}

The two components of V380\,Cyg have very different masses and radii, and thus evolutionary 
stages, so are a very good test of theoretical stellar models. The accuracies of the mass and 
radius measurements is generally 2\%, limited by the quality of the available light curves 
and the faintness of the secondary star in our spectra. A comparison with models is helped 
by our abundance analysis. Of particular interest is the effect of rotation on the evolution 
of massive stars, for which predictions are now available from Heger, Langer \& Woosley (2000), 
Heger \& Langer (2000) and Meynet \& Maeder (2000). These authors found that it makes a star 
of a given mass more luminous and longer-lived. The models also predict changes in the surface
 chemical composition in the sense that He and N become enriched whilst C and O are depleted.
 These changes are stronger for higher masses and/or initial rotational velocities.

To our knowledge, the only study which compares rotational evolutionary models with empirical 
data on high-mass stars in close binaries is that of Hilditch (2004). There is a known 
discrepancy between masses inferred from the location in the HR diagram relative to the model 
evolution tracks, and dynamically determined masses (Herrero et al.\ 1992). Hilditch found that 
the discrepancy did not disappear when rotating stellar models were used. We begin our comparison 
against V380\,Cyg with the models of Schaller et al.\ (1992) and Claret (1995) for solar 
metallicity ($Z=0.02$), moderate overshooting ($\alpha_{\rm ov} = 0.2$) and no rotational 
effects. Both good and bad agreement is seen for the Schaller models (Fig.\,\ref{fig:plotshaller},
 left panels): star A is overluminous for its mass, whereas the predictions for star B perfectly
 match its position. From the isochrones (green dotted lines) we estimate the age of the stars 
to be $\tau = 9.5 \pm 0.5$\,Myr. A similar situation occurs for the Claret models 
(Fig.\,\ref{fig:plotshaller}, centre panels) for a metal abundance of $Z=0.02$, whilst the 
models for $Z=0.01$ fail to match both stars. In the $\Teff - \logg$ diagram the situation 
is the same: the models can match star B but strongly disagree with the properties of star A.

Fig.\,\ref{fig:plotrot} (right panels) compares evolutionary models including rotational 
effects with the observed properties of V380\,Cyg. We interpolated in the tables published 
by Geneva group\footnote{\tt http://obswww.unige.ch/Recherche/evol/tables\_rot\-zams/} 
(Ekstr\"om et al.\ 2008). The models are parametrised by the ratio of the rotational velocity 
to the critical velocity, $\Omega/\Omega_{\rm crit}$. The models for $\Omega/\Omega_{\rm crit} 
= 0.1$ (green tracks), and $0.5$ (blue tracks) both fail to match the luminosity of star A. 
The discrepancy amounts to about 1.5\Msun. Star B is also slightly above evolutionary tracks 
for its mass (dotted lines), but agrees within the errorbars. The $\Teff - \logg$ diagram 
reveals the same story. It is interesting to note that in rotational evolution models the 
position of star A is around a critical evolutionary stage. Since rotational mixing extends 
the MS lifetime of high-mass stars, depending on initial rotational velocity, star A is either 
in the hydrogen-shell-burning phase or just at the `blue hook' at the TAMS.

Further insight into V380\,Cyg's evolution comes from its photospheric chemical composition. 
We interpolated in the Ekstr\"{o}m et al.\ (2008) tables to the mass of star A. 
For $\Omega/\Omega_{\rm crit} = 0.10$ ($V_{\rm ini} \sim 40$ \kms) the increase in the 
photospheric helium abundance at the end of core H-burning is 0.5\%, whilst N/C $=$ 0.69 
and N/O $=$ 0.24. Increasing to $\Omega/\Omega_{\rm crit} = 0.30$ gives a helium enhancement 
of 6\%, and a further increase in N at the expense of C and O (N/C $=$ 1.12 and N/O $=$ 0.40). 
Our analysis of the helium abundance in star A's photosphere (Sec.\,\ref{sec:abuhe}), 
gives a solar value, and we concluded that no enrichment is observed. Also, abundances 
of CNO elements (Sec.\,\ref{sec:abucno}) give N/C $=$ 0.25 and N/O $=$ 0.11 which is 
exactly the initial (ZAMS) values for the models. The photospheric chemical composition 
appears to be untouched even though the star has reached the end of the core H-burning 
phase. Models for $\Omega/\Omega_{\rm crit} = 0.10$, 0.30 and 0.50 at TAMS would have 
rotational velocities of about 20, 80 and 160 \kms, respectively. Since star A's rotational 
velocity is now $\sim$100\kms\ we interpolated in the tables to find that at the TAMS 
a 13.13\Msun\ model predicts N/C $\sim$ 0.40 and N/O $\sim$ 0.14, while initial values 
for this model are (N/C)$_{\rm ini} \sim$ 0.30 and (N/O)$_{\rm ini}$ $\sim$ 0.11. 
The later ratio is exactly what we found for star A, whilst no changes are observed 
in N/C ratio. We conclude that no changes in the abundances of the CNO elements were 
detected, which is contrary to model predictions.

Why might this be? The evolution of stars in close binaries can be modified by several 
effects. In particular, tides affect their rotational velocities and so the efficiency 
of rotational mixing. This has been studied theoretically using detailed calculations 
by De Mink et al.\ (2009), but unfortunately only for masses larger than V380\,Cyg\,A. 
De Mink et al.\ found large N enhancements for the shortest-period binaries, for which 
rotational velocities are the highest, and for high-mass primary stars which are close 
to filling their Roche lobes. But, for the relatively long orbital period of V380\,Cyg, 
12.4\,d, surface abundance changes would be small (and probably too small for us to have 
detected). We hope that future models will extend to the parameter space occupied by 
V380\,Cyg, allowing a direct comparison between observation and theory.

%%%%%%%%%%%%%%%%%%%%%%%%%%%%%%%%%%%%%%%%%%%%%%%%%%%%%%%%%%%%%%%%%%%%%%%%%%%%%%%%%%%%%%%%%%%%%%%%%%%%%%%%%%%%%%%%%%

\section{Summary}

We have presented a study of the eccentric eclipsing and double-lined spectroscopic binary 
V380\,Cyg. The system is particularly interesting because it contains an evolved star which
 is at the TAMS, so is ideal for probing theoretical evolution models which include 
rotationally induced mixing.

New high-resolution and high-S/N spectra were secured at several telescopes, and a spectral 
disentangling technique was applied to obtain the individual spectra of the components, and 
to (simultaneously) derive the spectroscopic orbit. Although the secondary star contributes 
only a very small fraction of the system light (6\%), we were able to attain a much higher 
precision than previously achieved. Coupled with a reanalysis of the existing light curves of 
this system, we have determined the masses and radii of the two stars to accuracies of 1--2\%. 
The light ratios found in the light curve analysis also allowed the disentangled component spectra
 to be accurately renormalised to the continuum level. These spectra were used to derive the 
effective temperatures of the two stars using two independent methods, helped by the accurate 
surface gravity values obtained from their known masses and radii.

The spectrum of V380\,Cyg is dominated by the primary star, and its disentangled spectrum has 
a S/N approaching 1000 per pixel and a wavelength coverage of 3900--7000\,\AA. We performed 
a detailed abundance analysis by fitting non-LTE spectra to the observed line profiles. We 
found no abundance excesses for helium or metallic species, and in fact the abundances are 
those of a typical B star. The helium abundance analysis showed a broad range of values from 
different lines, which we attribute to the failure of calculations for the lines formed higher 
in the photosphere due to the asphericity of the star. This problem deserves to be investigated 
further. It also remains possible that helium stratification might be the cause.

We have compared the bulk properties of V380\,Cyg to theoretical calculations from stellar 
models both with and without rotational effects. No set of models fully satisfies the observations, 
as the primary is always overluminous for its measured mass. We are still faced with well-known 
discrepancy between masses inferred from evolutionary models and those which have been dynamically
 measured. The mass discrepancy amounts to about 1.5\Msun\ for models including rotational 
evolution, and about 2.5\Msun\ for models which include only core overshooting, even though 
we have revised the mass of V380\,Cyg\,A upwards by a substantial amount. The surface chemical 
composition of high-mass stars should be an important probe of theoretical models. Recent model 
calculations which account for rotationally induced mixing predict changes in the abundances of 
elements involved in the CNO cycle in high-mass stars: helium and nitrogen should be enriched and
 carbon and oxygen depleted (strongly depending on the mass and initial rotational velocity). 
For a 13\Msun\ star all available models predicted changes at the end of core hydrogen burning, 
but the abundance pattern we derived in this work resembles that of a star at the ZAMS. We can 
conclude only that a detailed analysis performed for V380\,Cyg\,A does not corroborate theoretical
 predictions of the chemical evolution of high-mass stars. We are currently engaged in extending
 the sample of close binaries for which such comparisons can be made.

%%%%%%%%%%%%%%%%%%%%%%%%%%%%%%%%%%%%%%%%%%%%%%%%%%%%%%%%%%%%%%%%%%%%%%%%%%%%%%%%%%%%%%%%%%%%%%%%%%%

\section*{Acknowledgements}

Based on observations collected at the Centro Astron\'omico Hispano Alem\'an (CAHA) at Calar Alto,
operated jointly by the Max-Planck Institut f\"ur Astronomie and the Instituto de Astrof\'{\i}sica
de Andalucía (CSIC), and on observations made with the Nordic Optical Telescope, operated on the
island of La Palma jointly by Denmark, Finland, Iceland, Norway, and Sweden, in the Spanish
Observatorio del Roque de los Muchachos of the Instituto de Astrofisica de Canarias.
We would like to thank several colleagues for obtaining spectra of V380\,Cyg at the 2 m telescope
in Ond\v{r}ejov: M.~\v{S}lechta, J.~Polster, M.~Ceniga, M.~Netolick\'{y}, P.~\v{S}koda, V.~Votruba, 
P.~Hadrava, D.~Kor\v{c}\'{a}kov\'{a},
J.~Libich, K.~Uytterhoeven, B.~Ku\v{c}erov\'{a} and J.~Kub\'{a}t.
We acknowledge a constructive and timely response fro our reviewer Prof.~Philip Dufton.
This work was funded through a research grant to KP from Croatian Ministery of Science and Education.
JS acknowledges financial support from STFC in the form of grant number ST/F002599/1.
This work was partly funded from a research grant to PK from ESA PECS project No.\,98058.

%%%%%%%%%%%%%%%%%%%%%%%%%%%%%%%%%%%%%%%%%%%%%%%%%%%%%%%%%%%%%%%%%%%%%%%%%%%%%%%%%%%%%%%%%%%%%%%%%%%

%%%%%%%%%%%%%%%%%%%%%%%%%%%%%%%%%%%%%%%%%%%%%%%%%%%%%%%%%%%%%%%%%%%%%%%%%%%%%%%%%%%%%%%%%%%%%%%%%%%

\appendix

\section{Observing logs for the data presented in this work}

The tables in this section contain observing logs for each of the fours sets of spectroscopic
observations used in this work. In each case the orbital phases have been calculated with
ephemeris derived by Guinan et al.\ (2000).

\begin{table} \centering
\caption{\label{tab:obs:ondrejov:red} Observing log for the
Ond\v{r}ejov red spectra of V380\,Cyg}
\begin{tabular}{cccrr} \hline
Set & ID       & HJD         & Phase      & S/N \\
\hline
AUO & nh060033 & 53224.5345  &  963.1629  & 406 \\
AUO & nj050015 & 53284.3204  &  967.9744  & 476 \\
AUO & nj110016 & 53290.3615  &  968.4607  & 523 \\
AUO & of190036 & 53541.4822  &  988.6701  & 256 \\
AUO & pi240037 & 54003.3681  & 1025.8419  & 327 \\
AUO & pj100025 & 54019.4183  & 1027.1337  & 302 \\
AUO & pj160020 & 54025.3728  & 1027.6128  & 315 \\
AUO & pj170017 & 54026.2822  & 1027.6860  & 329 \\
AUO & pj200028 & 54029.3709  & 1027.9347  & 363 \\
AUO & qd020024 & 54193.5319  & 1041.1460  & 433 \\
AUO & qd040052 & 54195.5864  & 1041.3114  & 240 \\
AUO & qd110012 & 54202.6237  & 1041.8779  & 282 \\
AUO & qd130029 & 54204.4341  & 1042.0234  & 234 \\
AUO & qd140035 & 54205.4133  & 1042.1023  & 329 \\
AUO & qd150023 & 54206.4880  & 1042.1888  & 302 \\
AUO & qd160023 & 54207.5143  & 1042.2715  & 320 \\
AUO & qd190051 & 54210.5260  & 1042.5138  & 306 \\
AUO & qd200076 & 54211.5794  & 1042.5984  & 154 \\
AUO & qd220035 & 54213.4753  & 1042.7512  & 147 \\
AUO & qd220036 & 54213.4893  & 1042.7522  & 424 \\
AUO & qd230013 & 54214.4278  & 1042.8279  & 377 \\
AUO & qd250015 & 54216.5346  & 1042.9973  & 645 \\
AUO & qd300010 & 54221.3884  & 1043.3878  & 365 \\
AUO & qe010020 & 54221.5111  & 1043.3978  & 153 \\
AUO & qe030013 & 54224.3838  & 1043.6289  & 472 \\
AUO & qe170014 & 54238.4871  & 1044.7640  & 185 \\
AUO & qf230025 & 54275.5113  & 1047.7437  & 231 \\
AUO & qg140028 & 54296.5112  & 1049.4337  & 268 \\
AUO & qg140030 & 54296.5256  & 1049.4349  & 304 \\
AUO & qg170012 & 54299.3604  & 1049.6628  & 371 \\
AUO & qg260022 & 54308.5021  & 1050.3988  & 342 \\
AUO & qh050031 & 54318.4729  & 1051.2010  & 313 \\
AUO & qh130008 & 54326.3361  & 1051.8339  & 191 \\
AUO & qh130010 & 54326.3563  & 1051.8354  & 396 \\
AUO & qh170015 & 54330.4312  & 1052.1633  & 180 \\
\hline \end{tabular} \end{table}

\begin{table} \centering
\caption{\label{tab:obs:ondrejov:blue} Observing log for the
Ond\v{r}ejov blue spectra of V380\,Cyg.}
\begin{tabular}{clcrcr} \hline
Set & ID & HJD & Phase & $\lambda_{\rm c}$ (\AA) & S/N \\
\hline
AUO & ni170011 & 53266.3564  &   966.5286  &  4500 & 231 \\
AUO & ni180017 & 53267.3871  &   966.6116  &  4500 & 379 \\
AUO & nj050010 & 53284.2734  &   967.9706  &  4500 & 269 \\
AUO & nj100002 & 53289.2876  &   968.3742  &  4200 &  95 \\
AUO & nj120017 & 53291.2617  &   968.5330  &  4500 &  67 \\
AUO & nj240036 & 53303.4095  &   969.5107  &  4500 & 253 \\
AUO & nj250003 & 53304.2597  &   969.5789  &  4500 & 304 \\
AUO & nk250003 & 53335.2304  &   972.0715  &  4500 & 264 \\
AUO & oe260012 & 53517.4736  &   986.7380  &  4350 & 224 \\
AUO & pj160016 & 54025.3308  &  1027.6096  &  4900 & 139 \\
AUO & qd160028 & 54207.5544  &  1042.2747  &  4900 & 213 \\
AUO & qd190080 & 54210.5981  &  1042.5195  &  4900 &  83 \\
AUO & qd220060 & 54213.5538  &  1042.7574  &  4900 & 210 \\
AUO & qd270032 & 54218.5272  &  1043.1577  &  4900 & 254 \\
AUO & qd280024 & 54219.4498  &  1043.2318  &  4900 & 187 \\
AUO & qd290038 & 54220.5512  &  1043.3206  &  4900 & 176 \\
AUO & qe030018 & 54224.4380  &  1043.6333  &  4900 & 191 \\
AUO & qe180020 & 54239.4456  &  1044.8411  &  4350 & 129 \\
AUO & qf090021 & 54261.5286  &  1046.6183  &  4440 & 150 \\
AUO & qf110020 & 54263.4758  &  1046.7751  &  4400 & 122 \\
AUO & qf160017 & 54268.3798  &  1047.1697  &  4400 &  76 \\
AUO & qg140025 & 54296.4497  &  1049.4287  &  4400 & 198 \\
AUO & qg160015 & 54298.5406  &  1049.5969  &  4400 & 148 \\
AUO & qh050018 & 54318.3959  &  1051.1948  &  4400 & 148 \\
\hline \end{tabular} \end{table}

\begin{table} \centering
\caption{\label{tab:obs:victoria} Observing log for the
Victoria red spectra of V380\,Cyg.}
\begin{tabular}{clcrcr} \hline
Set & ID       & HJD         & Phase & S/N \\
\hline
DAO & 3c05618  &  54232.9356 & 1044.3172 & 306 \\
DAO & 3c07248  &  54259.9013 & 1046.4873 & 287 \\
DAO & 3c07992  &  54274.8548 & 1047.6907 & 248 \\
DAO & 3c08291  &  54276.9189 & 1047.8569 &  48 \\
DAO & 3c12337  &  54340.7187 & 1052.9914 & 236 \\
\hline \end{tabular} \end{table}

\begin{table} \centering
\caption{\label{tab:obs:int} Observing log of
the NOT and CAHA spectra of V380\,Cyg.}
\begin{tabular}{rcccr} \hline
Set  & ID   & HJD          & Phase      & S/N  \\
\hline
NOT  &      & 54039.37711  & 1028.7399  & 166  \\
NOT  &      & 54039.38248  & 1028.7403  & 309  \\
NOT  &      & 54039.39750  & 1028.7416  & 242  \\
CAHA & A167 & 54607.46305  & 1074.4586  &  56  \\
CAHA & A168 & 54607.46750  & 1074.4590  & 146  \\
CAHA & A181 & 54607.61685  & 1074.4709  & 156  \\
CAHA & A182 & 54607.62454  & 1074.4716  & 176  \\
CAHA & A190 & 54607.67179  & 1074.4753  & 144  \\
CAHA & A191 & 54607.67839  & 1074.4760  & 130  \\
CAHA & B161 & 54608.63799  & 1074.5530  & 116  \\
CAHA & B162 & 54608.64601  & 1074.5536  & 136  \\
CAHA & B164 & 54608.66251  & 1074.5552  &  80  \\
CAHA & B165 & 54608.67026  & 1074.5558  & 164  \\
CAHA & C181 & 54609.55296  & 1074.6268  &  32  \\
CAHA & C182 & 54609.56075  & 1074.6274  &  20  \\
CAHA & D158 & 54610.52706  & 1074.7052  & 144  \\
CAHA & D159 & 54610.53474  & 1074.7058  & 148  \\
CAHA & H200 & 54691.47874  & 1081.2201  &  66  \\
CAHA & H201 & 54691.48664  & 1081.2207  &  54  \\
CAHA & H202 & 54691.49426  & 1081.2213  &  60  \\
CAHA & H220 & 54691.67199  & 1081.2355  &  42  \\
CAHA & H221 & 54691.67967  & 1081.2361  &  36  \\
CAHA & H222 & 54691.68733  & 1081.2368  &  36  \\
CAHA & I191 & 54692.33182  & 1081.2887  &  60  \\
CAHA & I192 & 54692.33942  & 1081.2893  &  64  \\
CAHA & I219 & 54692.60461  & 1081.3107  &  70  \\
CAHA & I220 & 54692.61226  & 1081.3113  &  70  \\
CAHA & I221 & 54692.61994  & 1081.3119  &  84  \\
CAHA & J215 & 54693.61367  & 1081.3917  & 124  \\
CAHA & J216 & 54693.62132  & 1081.3923  & 120  \\
CAHA & J217 & 54693.62911  & 1081.3931  &  72  \\
CAHA & K189 & 54694.31946  & 1081.4486  & 156  \\
CAHA & K190 & 54694.32706  & 1081.4493  & 152  \\
CAHA & K221 & 54694.59423  & 1081.4707  &  96  \\
CAHA & K222 & 54694.59957  & 1081.4709  & 130  \\
CAHA & K223 & 54694.60717  & 1081.4716  & 118  \\
CAHA & M184 & 54696.32683  & 1081.6102  & 232  \\
CAHA & M208 & 54696.53837  & 1081.6272  & 144  \\
CAHA & M209 & 54696.54484  & 1081.6276  & 126  \\
CAHA & M210 & 54696.55136  & 1081.6282  & 120  \\
CAHA & M217 & 54696.58198  & 1081.6306  & 154  \\
CAHA & M218 & 54696.58974  & 1081.6313  & 144  \\
CAHA & M219 & 54696.59739  & 1081.6328  & 108  \\
\hline \end{tabular} \end{table}

%%%%%%%%%%%%%%%%%%%%%%%%%%%%%%%%%%%%%%%%%%%%%%%%%%%%%%%%%%%%%%%%%%%%%%%%%%%%%%%%%%%%%%%%%%%%%%%%%%%
\label{lastpage} \end{document}